\documentclass[fleqn,usenatbib]{rasti}
\usepackage{newtxtext,newtxmath}

\usepackage[T1]{fontenc}

\DeclareRobustCommand{\VAN}[3]{#2}
\let\VANthebibliography\thebibliography
\def\thebibliography{\DeclareRobustCommand{\VAN}[3]{##3}\VANthebibliography}

\usepackage{graphicx}	
\usepackage{amsmath}
\usepackage{siunitx}
\usepackage{xcolor}
\usepackage[dvipsnames]{xcolor}
\usepackage{booktabs}
\usepackage{textcomp}
\usepackage{multirow}
\usepackage{orcidlink}

\newcommand{\profe}{\texttt{PROFE}}

\title[Exoplanet Transits with OPTICAM]{Data reduction method for OPTICAM multiband time series of transiting exoplanets}

\author[S. P\'aez et al.]{
S. P\'aez$^{1}$\thanks{E-mail: spaez@astro.unam.mx}\orcidlink{0000-0001-6381-7120},
Y. G\'omez Maqueo Chew$^{1}$\orcidlink{0000-0002-7486-6726},
L. H. Hebb$^{2}$
\\
$^{1}$Universidad Nacional Aut\'onoma de M\'exico, Instituto de Astronom\'ia, AP 70-264, 04510. Ciudad de M\'exico, M\'exico.\\
$^{2}$Department of Physics, Hobart and William Smith Colleges, Geneva, New York, 14456, USA\\
}

\date{Accepted 2026 March 07. Received 2026 February 27; in original form 2025 September 09}

\pubyear{\the\year{}}

\begin{document}
\label{firstpage}
\pagerange{\pageref{firstpage}--\pageref{lastpage}}
\maketitle

\begin{abstract}

We present a methodology for acquiring and reducing transiting exoplanet light curves obtained with the OPTICAM instrument in the Observatorio Astron\'omico Nacional en la Sierra de San Pedro M\'artir (OAN-SPM). The OPTICAM sCMOS detectors generate significant warm pixels at exposures $\geq$10s, affecting both science and calibration frames. These warm pixels are not removed by standard dark subtraction because they vary unpredictably from frame to frame. We evaluate six pre-processing methods applied to science and calibration images using the transit of TOI-7149~b observed in g$^\prime$r$^\prime$i$^\prime$. A median filter with a 3$\times$3-pixel window minimizes the effect of warm pixels without affecting stellar signals. This median filter best reduces dispersion and red noise in the light curves when stellar peak counts are close to the dark current level. The improvement is less significant when the stellar peak is several thousand counts above the dark current level. We fit a multiband transit model to the light curves, measuring photometric precision, correlated noise, and retrieved planetary parameters. The transit model fitted to the light curves with pre-processing using a 3$\times$3-median filter achieves the highest Bayesian evidence. Thus, it is our recommended method for correcting warm pixels. Finally, we present a reduction pipeline that combines Python modules (\profe) and AstroImageJ to implement our proposed method for OAN-SPM 2.1m+OPTICAM transiting planet observations.
\end{abstract}




\section{Introduction}

Simultaneous multiband photometry enhances our understanding of exoplanetary systems by capturing wavelength-dependent variations in the transit light curve. Eclipsing binary stars may mimic planetary transits yet exhibit chromatic transit depths due to their intrinsic brightness, allowing multiband observations to identify these false positives and validate the planetary nature of a signal \citep[e.g.,][]{evans2010anapriori,mignon2023characterisation}. Similarly, comparing transit data across multiple bands enables the characterization of stellar activity features, such as spots \citep[e.g.,][]{rackman2018thetransit} and faculae, that could affect the determination of the exoplanet properties \citep[e.g.,][]{rosich2020correcting}.  
Although the transit depth may vary as a function of wavelength because of the planet's atmosphere  \citep[i.e., transmission spectroscopy; e.g.,][]{narita2015muscat}, these observations are typically carried out from space-based telescopes \citep[e.g.,][]{rustamkulov2023JWST,fisher2024JWST/NIRISS} because ground-based, broadband observations do not have sufficient precision to characterize the atmospheres. 

Precise, multiband transit photometry is not trivial. Differential photometry, a commonly used technique applied for each filter, can achieve high precision even under suboptimal conditions \citep{hartley2023optimized}. However, it remains susceptible to noise from atmospheric effects \citep[e.g.,][]{osborn2015atmospheric}, instrumental systematics \citep[e.g.,][]{pont2006rednoise}, guiding drift, intrinsic stellar variability \citep[e.g.,][]{oshagh2018noise}, and the detectors themselves, whose manufacturing, environment, and operating parameters set the noise floor \citep[e.g.,][]{shao2024theimpact}.

Here we present the case of OPTICAM, a high-temporal-resolution optical instrument designed for use with the 2.1m telescope at the Observatorio Astron\'omico Nacional en la Sierra de San Pedro M\'artir OAN-SPM) in Baja California, M\'exico. The OPTICAM system employs a dual-beam splitter, enabling simultaneous observations in three independent channels. These channels can be configured with three of the SDSS filters: u$^\prime$ or g$^\prime$ for Channel 1; r$^\prime$ for Channel 2; and i$^\prime$ or z$^\prime$ for Channel 3. OPTICAM was originally designed to study rapidly varying astrophysical phenomena on timescales from milliseconds to minutes, including transient events in X-ray binaries, black hole and neutron star galactic transients, and accretion onto compact objects. Mounted on the OAN-SPM~2.1m telescope, it is optimized for photometric observations for targets ranging from 8 to 18 mag. For example, observations of an M3.5 dwarf (V=10.26 mag) using 0.1s exposures in u$^\prime$ band achieved a $\sim$20 mmag precision \citep[][their Fig.~12]{castro2024first}. In their Fig.~13, while monitoring a magnetic cataclysmic variable, observations in g$^\prime$r$^\prime$i$^\prime$ bands with 2s exposures yielded typical errors of 0.3\% for g$^\prime$ and r$^\prime$ bands and 0.1\% for i$^\prime$. 
These results demonstrate OPTICAM's capacity for high-cadence observations of high-energy astrophysical transient phenomena. 
Although not optimized explicitly for transiting exoplanet light curves, we demonstrate in this work the ability to acquire high-quality time-series observations of transiting exoplanets with OAN-SPM 2.1+OPTICAM.

A key feature of OPTICAM is its use of three scientific complementary meta-oxide-semiconductor (sCMOS) detectors. Historically, CCDs dominated astronomical observations \citep[e.g.][]{lesser2015asummary}, but modern sCMOS offer shorter readout times, lower read noise, lower power consumption, and lower cost \citep[e.g.,][]{greffe2022characterisation, liu2022research, khandelwal2024beyond}, making them well suited to high-cadence applications such as OPTICAM. Addressing the diverse noise sources requires careful attention to both data acquisition and reduction. OPTICAM sCMOS detectors produce significant warm pixels, a problem already reported in the literature \citep[e.g.,][]{Alarcon2023CMOS}. These warm pixels introduce red noise in the light curves. They cannot be corrected by standard reduction procedures, such as dark-frame subtraction and flat-field division, because they vary unpredictably from frame to frame.

In this context, we present a method for acquiring and reducing time-series simultaneous multiband observations of transiting exoplanets with OAN-SPM~2.1m+OPTICAM, especially addressing the warm-pixel problem. We characterize and describe in detail the behavior of the OPTICAM sCMOS detectors, the warm pixels they produce, and our observation strategies in Section~\ref{sec: characterization and plan}. In Section~\ref{sec: data reduction approaches}, we evaluate Gaussian convolution kernels and median filters as pre-processing methods for science and calibration images in the context of the TOI-7149 b transit light curve. We describe our transit fitting setup in Section~\ref{sec:juliet}. Finally, in Section~\ref{sec: proposed method} we identify the best reduction strategy and introduce the reduction pipeline that integrates \profe, our open-source Python code, and AstroImageJ \citep[AIJ;][]{collins2017astroimagej} to reduce these datasets.

\section{Frames characterization and observations plan} \label{sec: characterization and plan}

\subsection{Behavior of OPTICAM sCMOS detectors at long exposure times} \label{sec:detec. behavior}

OPTICAM incorporates three state-of-the-art Andor Zyla 4.3-Plus USB 3.0 sCMOS cameras with 2048$\times$2048 pixels CIS2020 detectors manufactured by Fairchild Imaging with a plate scale of 0.139\arcsec/pix in Channel 1, 0.140\arcsec/pix in Channel 2, and 0.166\arcsec/pix in Channel 3. These detectors support readout speeds of up to 40 frames per second in 16-bit mode, ensuring rapid and efficient data acquisition \citep{castro2024first}. 

\begin{figure*}\centering
  \includegraphics[width=17.8cm,trim={3.6cm 15.25cm 4.4cm 5.3cm},clip]{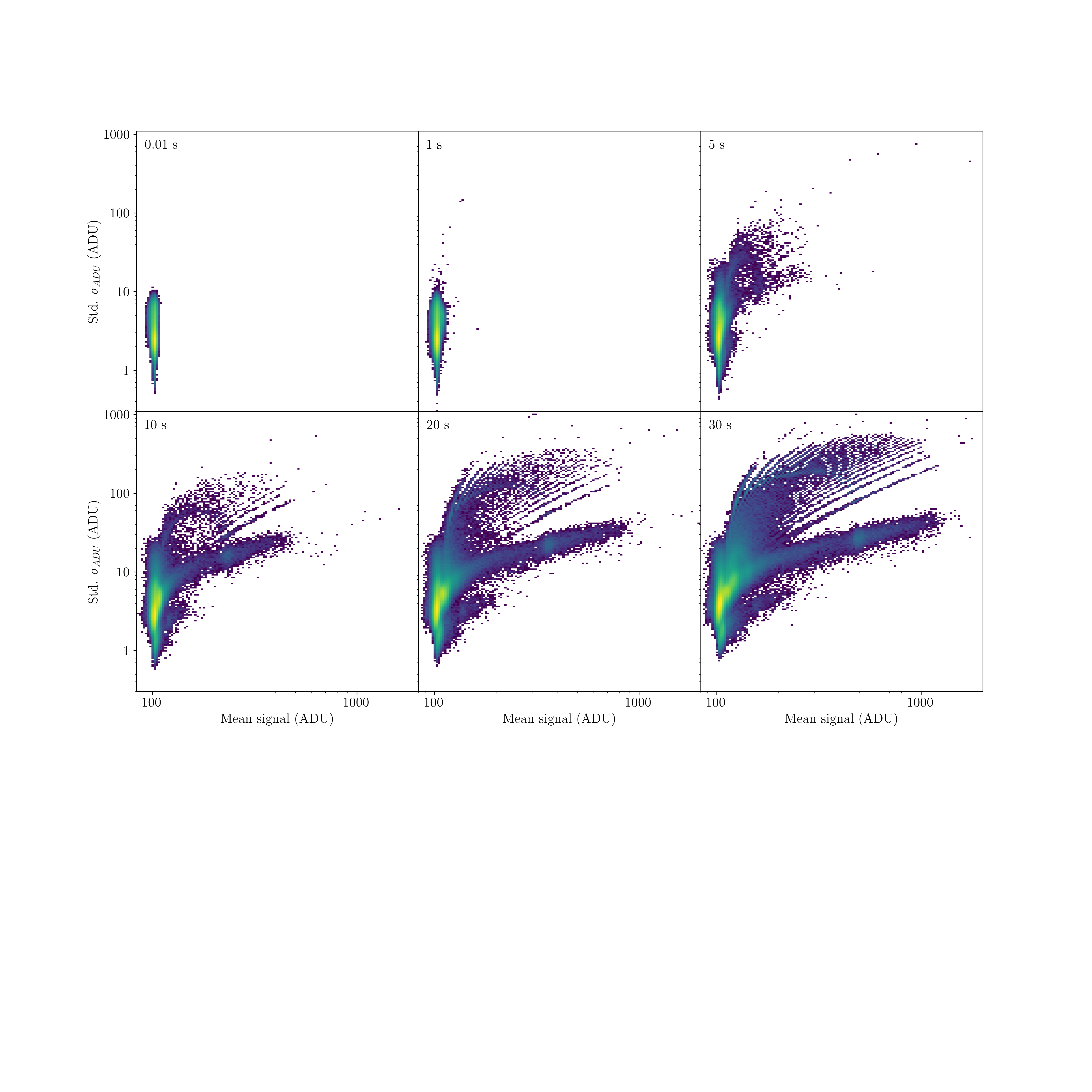}
  \caption{
  2D histogram of the standard deviation vs mean counts for each pixel in sequences of 25 darks with different exposure times (i.e., 0.01, 1, 5, 10, 20, and 30 seconds). For simplicity, all sequences are shown for OPTICAM's Channel 1, but Channels 2 and 3 behave similarly. The x- and y-axes are the same in each panel, and are chosen to optimize the comparison between the distributions with varying exposure time. A few pixels extend beyond the shown region for long exposures. As exposure time increases, the warm-pixel branches become progressively more structured, highlighting that both the average dark current and its frame-to-frame variability increase with integration time.}
  \label{fig:histograms}
\end{figure*}

\begin{figure}
    \includegraphics[width=\columnwidth,trim={0.95cm 0.6cm 1.55cm 0.8cm},clip]{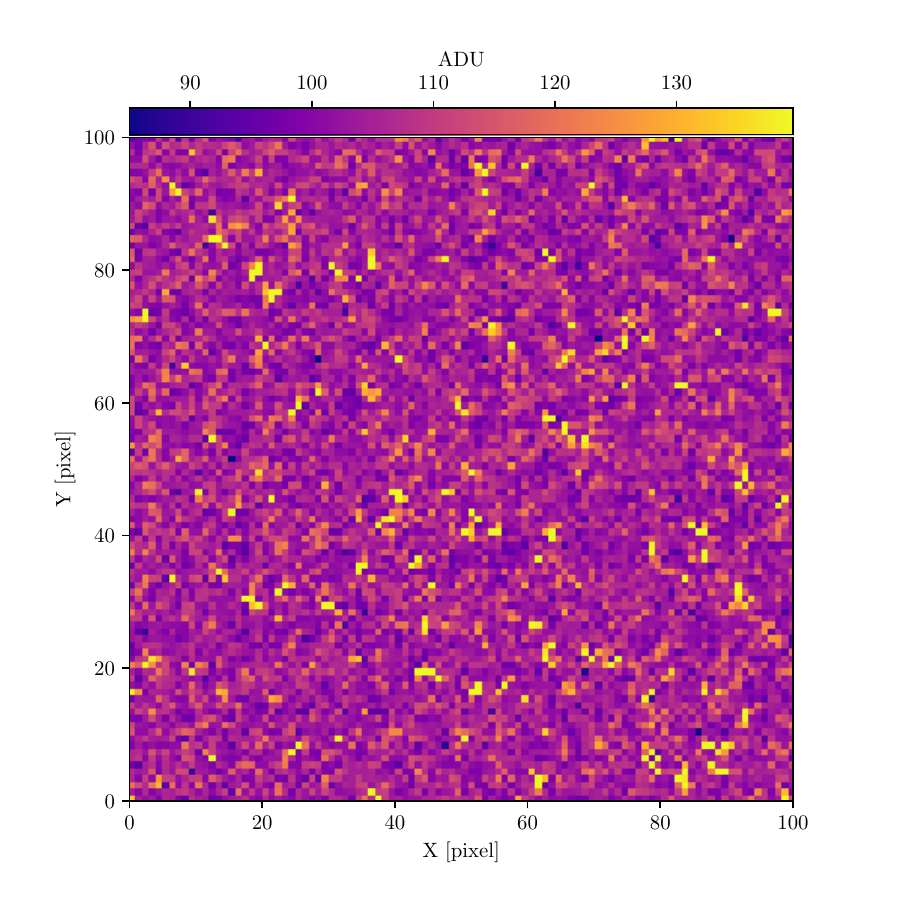}
  \caption{Spatial distribution of warm pixels in a $100\times100$-pixel subframe from a single dark with 30~s exposure time, where warm pixels appear scattered across the subframe. This subframe is in Channel 1, and in the other Channels the distribution of warm pixels is similar.} 
  \label{fig:raw_subframe_hist}
\end{figure}

To characterize the OPTICAM images, we followed the procedure detailed in \citet{Alarcon2023CMOS}. We used sequences of 25 darks each, acquired with different exposure times (i.e., 0.1, 1, 5, 10, 20 and 30~s) 
to calculate the standard deviation and the mean counts for each pixel in the sequence and show them as 2D histograms in  Fig.~\ref{fig:histograms}. In this figure,
at the shortest exposure time, pixels show minimal dispersion in both axes (i.e., mean $<$110 ADU and standard deviation $<$10 ADU). As the exposure time increases, the 2D histogram becomes broader by reaching larger mean ADU counts and/or a larger standard deviation.  The increase of the mean toward higher values can be due to the dark current or an increase in the pixel bias level, whereas the increase in the standard deviation is due to the Salt \& Pepper effect or an increased readout noise \citep[e.g.,][]{Alarcon2023CMOS}. 
We identify warm pixels in a sequence of dark frames as those having a mean counts above 110 ADU and/or having a standard deviation of more than 10 ADU. Figure~\ref{fig:raw_subframe_hist} shows that the warm pixels (bright in the subframe) are distributed throughout the full frame, and the standard deviation shows that their count value changes significantly in the sequence. 
Given that these 2D histograms are calculated from a sequence of 25 darks, where there is no stellar signal, we attribute this warm-pixel behavior of increased standard deviation in the counts and/or increase in the mean to an instrumental nature.

 A set of extended structures or ``branches'' composed of warm pixels becomes prominent in the 2D histograms for exposure times longer than 10~s, with the number of warm pixels also growing with exposure time. We quantify this fraction of warm pixels for seven distinct sequences of 25 darks with 30~s exposure time acquired between 2024 and 2025. 
 For Channel 1, we find that the warm pixels represent between $\sim$17--20\% of all pixels in the image, and 15\% of those remain warm throughout all sequences. For Channel 2, warm pixels make up between $\sim$30-35\% of all pixels, of which 27\% remain warm in all sequences. Similarly, for Channel 3, the warm pixels represent between $\sim 23-60$\%, of which 21\% remain warm throughout all sequences. 

\begin{figure*}
  \centering
  \includegraphics[width=17.8cm,trim={0.3cm 0.35cm 0.3cm 0.38cm},clip]{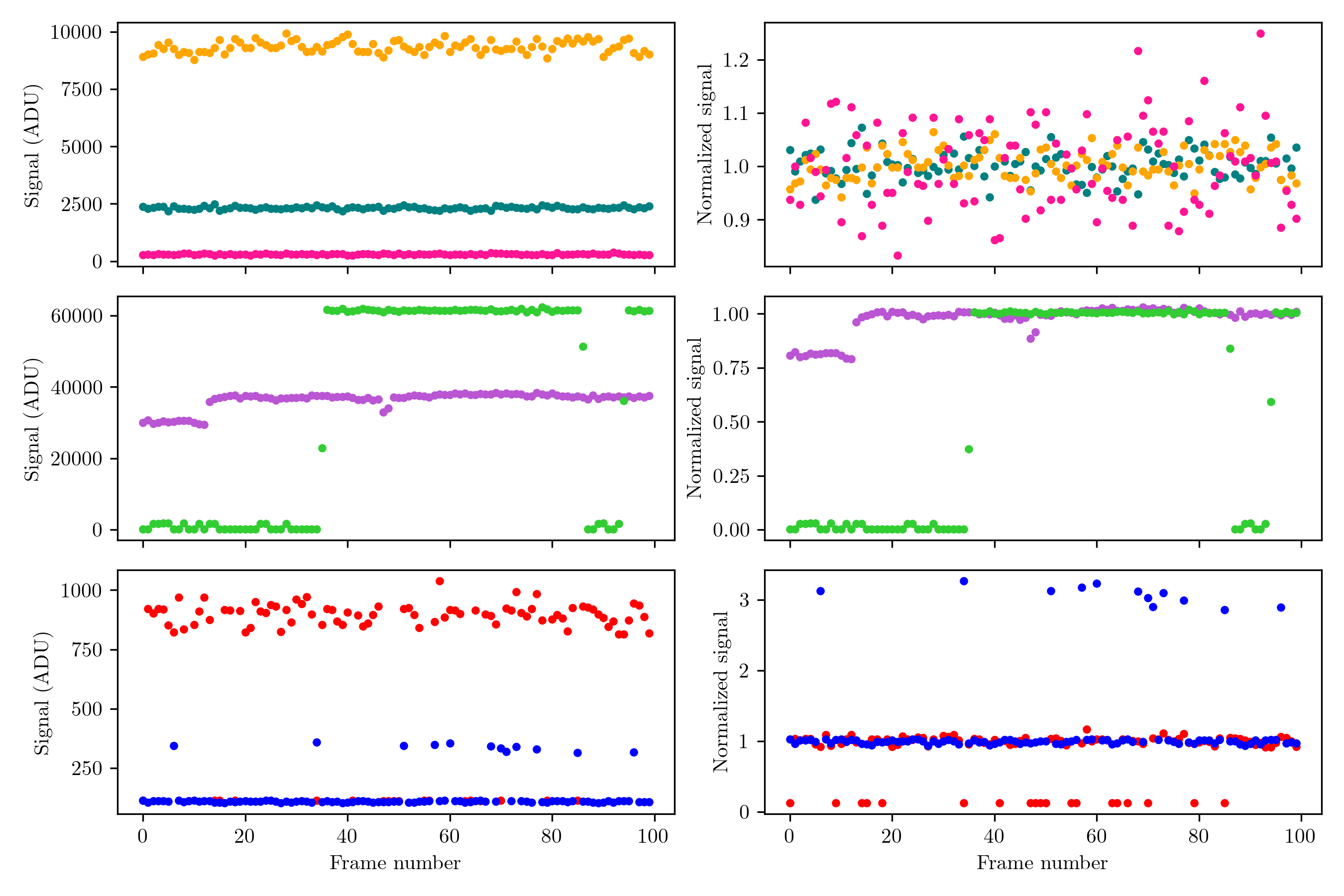}
  \caption{Examples of different types of warm pixels. We show the measured counts (left) and median-normalized counts (right) for distinct warm pixels as a function of consecutive frame number for 100 30-second darks. \textit{Top}: Type A warm pixels are dominated by an offset in threshold voltage, maintaining relatively constant high mean counts with about 10--20\% dispersion; the different colours correspond to different median levels (e.g., pink: 305 counts; teal: 2320 counts; orange: 9315 counts) but show the same behaviour. \textit{Middle}: Type B warm pixels exhibit unpredictable, stepwise changes in count level; the green example reaches nearly the detector's saturation limit and drops to lower values in other frames, while the purple example has an average of 37214 counts and shows a mixed behaviour between types A and B. \textit{Bottom}: Type C (blue) warm pixels remain near the median count value for most of the sequence but display occasional upward jumps (salt events), whereas Type D (red) warm pixels stay at high counts for most of the time series with intermittent drops to the median (pepper events). These different types of warm pixels trace the various branches seen in the 2D histograms and highlight the complexity of mitigating their effects in the light curves.}
  \label{fig:hotpixels lc}
\end{figure*}

We examined individual warm pixels to study their behavior in time. In Fig.~\ref{fig:hotpixels lc}, we show the raw (right) and normalized (left) counts as a function of frame number for the different types of warm pixels.
Guided by the 2D histograms shown in Fig.\ref{fig:histograms}, 
we categorize four types of warm-pixel behavior:
Type A is a warm pixel dominated by a higher bias level, 
which maintains elevated mean counts with a dispersion of $\sim$10-20\% throughout the time series (Fig.~\ref{fig:hotpixels lc}-top panels). Its standard deviation increases with the mean signal;
Type B is  a warm pixel that exhibits a stepwise behavior that is unpredictable and, in most cases, it reaches above 20\,000 ADU (Fig.~\ref{fig:hotpixels lc}-middle panels); 
Type C is a warm pixel that, for most of the sequence, has a count value that is similar to pixels that are not warm in the image, with occasional upward jumps to higher values increasing its standard deviation (Fig.~\ref{fig:hotpixels lc}-bottom panels in blue);
Type D is a warm pixel that stays at high counts for most of the sequence, with intermittent drops to values near the counts of pixels that are not warm, increasing its standard deviation (Fig.~\ref{fig:hotpixels lc}-bottom panels in red).
We identify that a given pixel may behave as a warm pixel in one time series, and not in another. A single pixel may also change between these categories and/or exhibit combinations of the behaviours described above during different observing sequences. In general, we find that the warm pixels exhibit unpredictable behavior; thus, a pixel mask is insufficient to fully mitigate their impact on the light curves. 
Given the large fraction of warm pixels in an image (up to 60\%), we consider that implementing outlier rejection for the warm pixels using either a static or a dynamic pixel mask is not suitable for our data. In the following sections, we explore different strategies based on data reconstruction to mitigate the effects of warm pixels in the transit light curves acquired with OPTICAM.

\subsection{Observability window: pointing and guiding}
\begin{figure}
    \centering
    \includegraphics[width=\columnwidth,trim={0.2cm 0.2cm 1.4cm 1.4cm},clip]{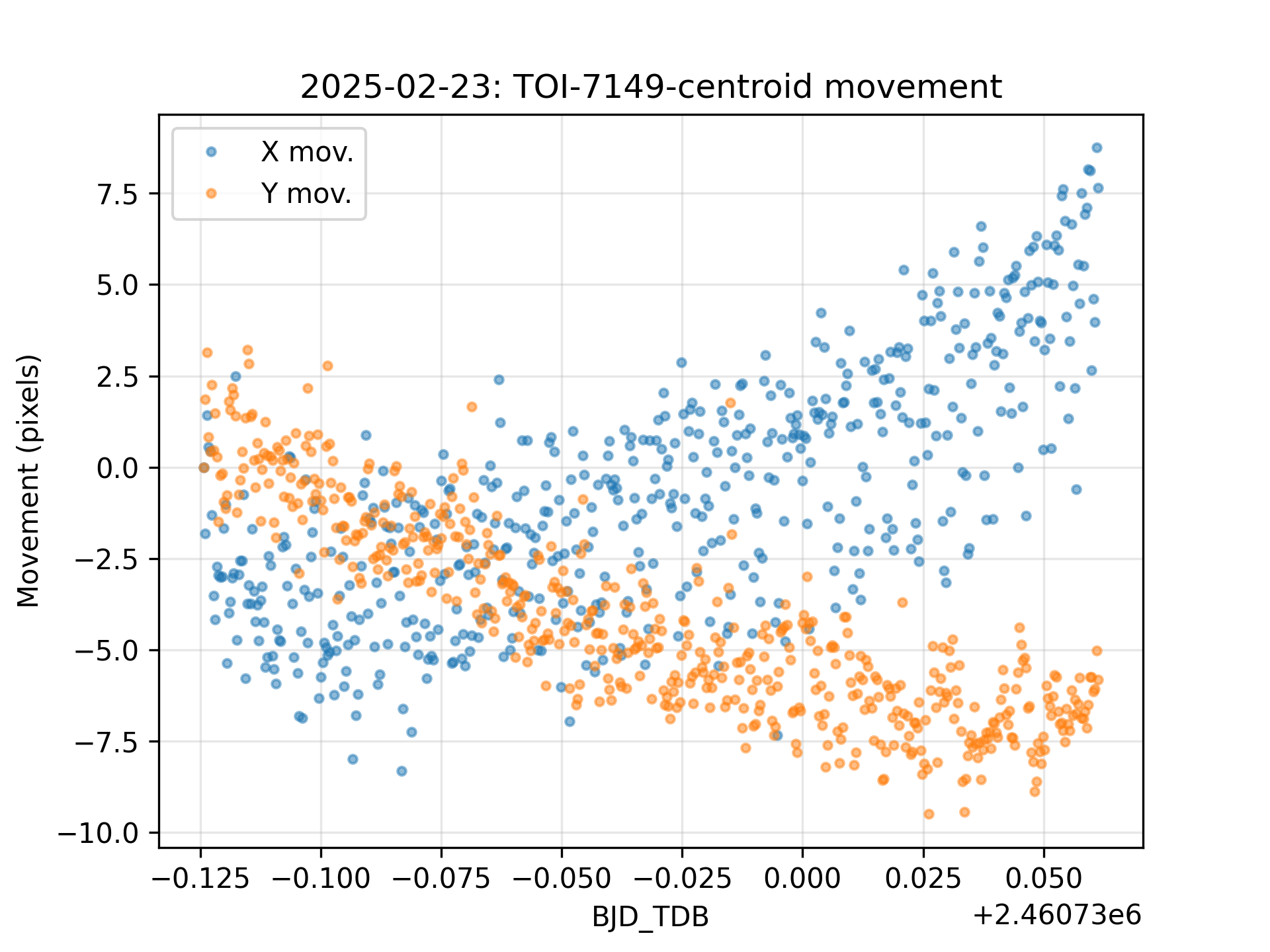}
    \caption{A target centroid movement over a 9.5-hour time series in channel 1. Blue dots are for X-axis movement, and orange dots are for Y-axis movement. Other channels show similar centroid movement patterns, with variations in the number of displaced pixels due to differences in pixel scale across channels.}
    \label{fig:guiding}
\end{figure}

The OAN-SPM~2.1m telescope features a horseshoe equatorial mount, which imposes a hard limit on hour angle of approximately $\pm$5 hours at declinations between $-40\degr$ and $+69\degr 40\arcmin$. As such, an essential part of planning the observations is to ensure that the transit window, including the out-of-transit phases needed for normalization, is observable within these telescope limits. We use the telescope's guiding system, choosing a guiding star with a magnitude between 6.5 and 9 mag and using exposure times between 1.5 and 4 seconds.  This setup keeps the star centroids within $\pm$30--40 pixels for time series lasting up to 9 hours. Figure~\ref{fig:guiding} shows a typical change in position of the centroid of the star over an entire time series in OPTICAM channel 1. Other channels show similar patterns, although the measured pixel displacement varies according to each channel's distinct pixel scale.   

\subsection{Observation campaigns and the transit of TOI-7149 b}\label{sec:observations}

In this paper, we present and analyse in detail a single OPTICAM transit observation of confirmed TESS planet TOI-7149~b \citep{Kanodia2025searching}, which we use to exemplify the reduction process, light-curve fitting, and to assess the robustness and performance of the proposed method. TOI-7149~b has a period  P$_{\rm orb}=2.6$ days and produces a transit with a depth of 12\% with a duration of $T_{14}=1.7256$ hours, and orbits an M dwarf with $T_{\rm eff} = 3363 \pm59$ K and magnitudes g$^\prime = 18.08$, r$^\prime = 16.87$, and i$^\prime = 15.409$ \citep{Kanodia2025searching}. The OPTICAM observations of this transit were acquired on 2025 February 26 (UT) with 30s exposures, yielding 534 science frames per filter that cover $\sim$4h, yielding an out-of-transit baseline of one hour before and one hour after the transit. 

On the night of the TOI-7149~b transit, we obtained 42 sky flats per filter with counts between 20,000 and 30,000 ADU counts to remain within the detectors' linear regime and to minimize Poisson noise \citep{howell2000handbook}. For this dataset, we acquired 25 dark frames with 30~s exposure time. Because OPTICAM does not have shutters, darks have to be scheduled during the nights, and in this case, we acquired the darks during the previous night to the observation. Bias frames are unnecessary because the sCMOS detectors’ architecture allows each pixel to process its signal independently with its own amplifier, and any remaining bias signal is accounted for when subtracting the darks.

Given that the OPTICAM's acquisition software synchronizes the start of each channel's exposure, the cadence in the light curve is given by the channel with the longest integration. We therefore used the same exposure time (30s) for all channels to optimize the observation time and minimize dead time.
OPTICAM provides a field of view of approximately 4.77\arcmin, 4.80\arcmin, and 5.67\arcmin\ for each channel, respectively \citep{castro2024first}. For that reason, we optimized the pointing of the observation to maximise the number of available comparison stars in the smallest field of view. It is necessary to ensure that at least one comparison star is present in all channels to perform differential aperture photometry.

\begin{table}
\caption{OPTICAM light-curve data for TOI-7149.}\label{tab:lcs}
\begin{tabular}{c c c c c c c}
\hline\hline\\
 BJD$_{\rm TDB}$      & Norm. Flux      & Flux Error      & Filter & Method  \\
 $- 2\,460\,000$ &  & 
 & \phantom{telescope} &\phantom{Filter} \\
\hline\\
729.875696     &        1.010        & 0.011 & g$^\prime$ & \texttt{st}\\
729.876043     &        1.032        & 0.011 & g$^\prime$ & \texttt{st}\\
... & ... & ... & ... \\
\hline\\
\end{tabular} \\
{\footnotesize The full table is available as online material.}
\end{table}

\section{Data reduction approaches} \label{sec: data reduction approaches}
To minimize the effect of the warm pixels, we tested six different reduction approaches. Each is detailed in the following sections. We compared the standard reduction (\texttt{st}; Section~\ref{sec:st reduction}) to reductions done after the pre-processing of the images with Gaussian convolution with kernel sizes of $\sigma$=1 pixel (\texttt{g1}) and $\sigma$=3 pixels (\texttt{g3}) in  Section~\ref{sec:gau_kernels}, and with median filters in Section~\ref{sec: median filter} with pixel-window sizes of 3$\times$3 pixels (\texttt{w3}), 5$\times$5 pixels (\texttt{w5}), and 7$\times$7 pixels (\texttt{w7}). The goal was to identify a method that effectively suppresses the warm-pixel contribution while maintaining the integrity of the astronomical signal, ensuring that key features of the data remain unaffected. In contrast to typical cosmic-ray rejection, where impacted pixels are first identified, and only those pixels are replaced, our Gaussian-convolution and median-filter pre-processing steps are applied globally to every pixel in both the science and calibration frames. 

We use AIJ\footnote{\href{https://www.astro.louisville.edu/software/astroimagej/}{www.astro.louisville.edu/software/astroimagej/}} (version 5.5.1) to perform differential photometry with calibrated time series for each reduction approach to obtain the TOI-7149~b transit light curves and fit each of those multiband light curves with the \texttt{juliet} package \citep[Section~\ref{sec:juliet};][]{Espinoza2019juliet} incorporating Gaussian processes (GPs) in each band to model the transit, and the normalization and correlated noise. With this, we obtain the planet parameters and the fit residuals for each reduction approach.

For the differential photometry of TOI-7149, we identified 19 potential comparison stars within the field of view. We selected the appropriate set of comparison stars for each filter by minimizing the RMS of the out-of-transit light curve obtained via the standard reduction. This process resulted in an ensemble of 15 stars for g$^\prime$, 10 stars for r$^\prime$, and 16 stars for i$^\prime$. To strictly evaluate the performance of the warm-pixel removal approaches, we used this set of comparison stars across all six reductions. This ensures that any variations in photometric precision or Bayesian evidence are solely attributable to the pre-processing method and not to changes in the comparison stars.

We decided against using PSF photometry because OPTICAM has two dichroics to enable the simultaneous three-band photometry. These optical elements produce channel-dependent, non-analytical PSFs, as shown in the isocontours of the stars (Sec.~\ref{sec:st reduction} to \ref{sec: median filter}). Given the complex optics of OPTICAM, it is not possible to perfectly focus all three channels simultaneously. Furthermore, the $\lesssim$5 arcmin fields usually lack a sufficient density of isolated stars to build a reliable spatially varying empirical PSF, and rapid seeing fluctuations during the time series may distort the PSF further, making fits unstable and prone to more correlated noise \citep{Mann20211ground}.

In Table~\ref{tab: metrics}, we summarise the metrics computed from the residuals of the fitted light curves for each reduction. For every band we report 
(a) the Bayesian evidence $\ln{\mathcal{Z}}$, computed directly by \texttt{juliet}, to assess the statistically preferred model for all three light curves simultaneously,
(b) the photometric precision of the light curves, i.e., the root-mean-square (RMS) scatter of the residuals, 
(c) the 10-min binned RMS, RMS$_{10\ \rm min}$ which estimate the  correlated noise,
(d) the median photometric errors (Median $\sigma$), 
(e) the ratio RMS/RMS$_{10\ \rm min}$, and 
(f) the reduced chi-squared, $\chi_\nu^2$.
The residuals are calculated between the data and the transit model, excluding the contribution of the GPs. Such that the correlated noise that is fitted by the GPs is kept for the quantification of the RMS and RMS$_{10\ \rm min}$.
These quantities allow us to compare the reductions in both per-band photometric performance and overall model preference. 
The Bayesian evidence quantifies the probability of the data given a model, marginalised over the model parameters. As discussed by \citet{Espinoza2019juliet}, differences in $\ln\mathcal{Z}$ provide a way to compare models with different levels of complexity or different noise treatments: models with larger $\ln Z$ are statistically preferred, with differences larger than $\Delta \mathcal{Z} > 5$ indicating strong evidence for a fitting over others.
In this work, we adopt $\ln{\mathcal{Z}}$ as the primary criterion to rank the reduction approaches, and we use the remaining metrics to diagnose how each reduction changes the noise in the light curves. Of those other metrics, we emphasize on RMS$_{\rm 10\ min}$ because warm pixels impact the light curves through time-correlated noise. Median $\sigma$, RMS/RMS$_{10\ \rm min}$, and $\chi_{\nu}^2$ are included as consistency checks on the phometric uncertainties and the fitting quality.

\begin{figure}
    \includegraphics[width=\columnwidth, trim={0.1cm 1.2cm 0.2cm 0.5cm},clip]{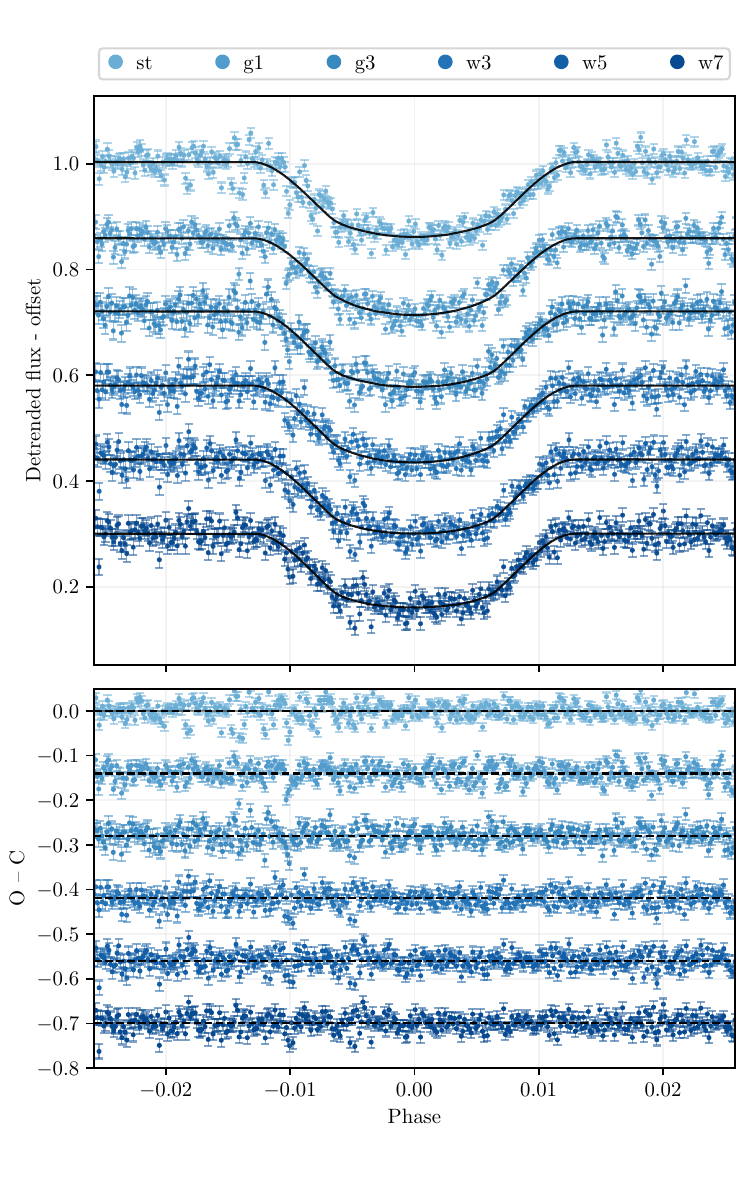}
    \caption{OPTICAM light curves, models, and residuals of TOI-7149 for different reductions in the g$^\prime$ filter. From top to bottom: standard reduction (\texttt{st}), Gaussian convolution kernel with 1-$\sigma$ (\texttt{g1}), Gaussian convolution kernel with 3-$\sigma$ (\texttt{g3}), median filter with 3$\times$3 pixel window size (\texttt{w3}), median filter with 5$\times$5 window size (\texttt{w5}), and median filter with 7$\times$7 window size (\texttt{w7}). \textit{Top:} Light curves (blue color scale) and models (black) with vertical offset. \textit{Bottom:} Residuals (blue color scale) for each reduction with vertical offset.    These different light curves, models, and residuals from various reductions of the same data demonstrate the impact of the reduction approach on the final results. The g$^\prime$ filter light curves are the most dispersed because in this filter, the target star achieves fewer counts than in the r$^\prime$ and i$^\prime$ filters and therefore is closer to the dark current level of the image.} 
    \label{fig: gp_lc}
\end{figure}

\begin{figure}
    \includegraphics[width=\columnwidth, trim={0.1cm 1.2cm 0.2cm 0.5cm},clip]{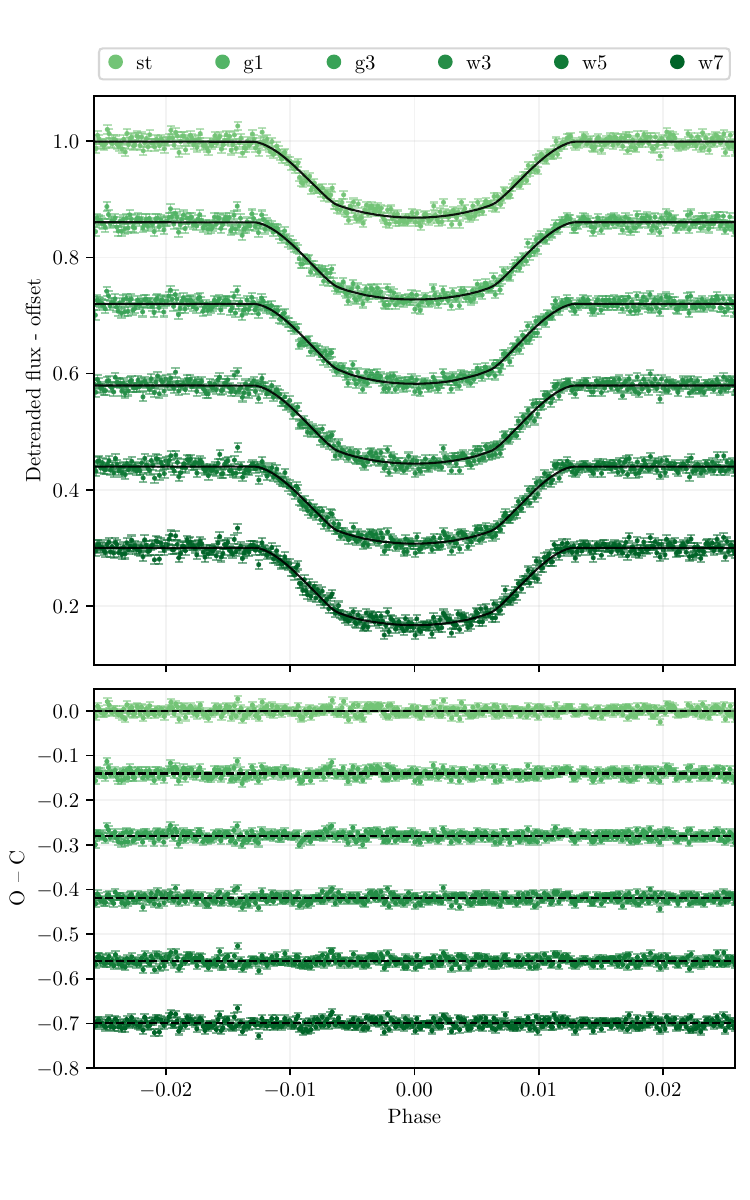}
    \caption{Similar to Fig.~\ref{fig: gp_lc} for r$^\prime$ filter. The r$^\prime$ filter light curves are less dispersed than in the g$^\prime$ filter, but more dispersed than in the i$^\prime$ filter.} 
    \label{fig: rp_lc}
\end{figure}

\begin{figure}
    \includegraphics[width=\columnwidth, trim={0.1cm 1.2cm 0.2cm 0.5cm},clip]{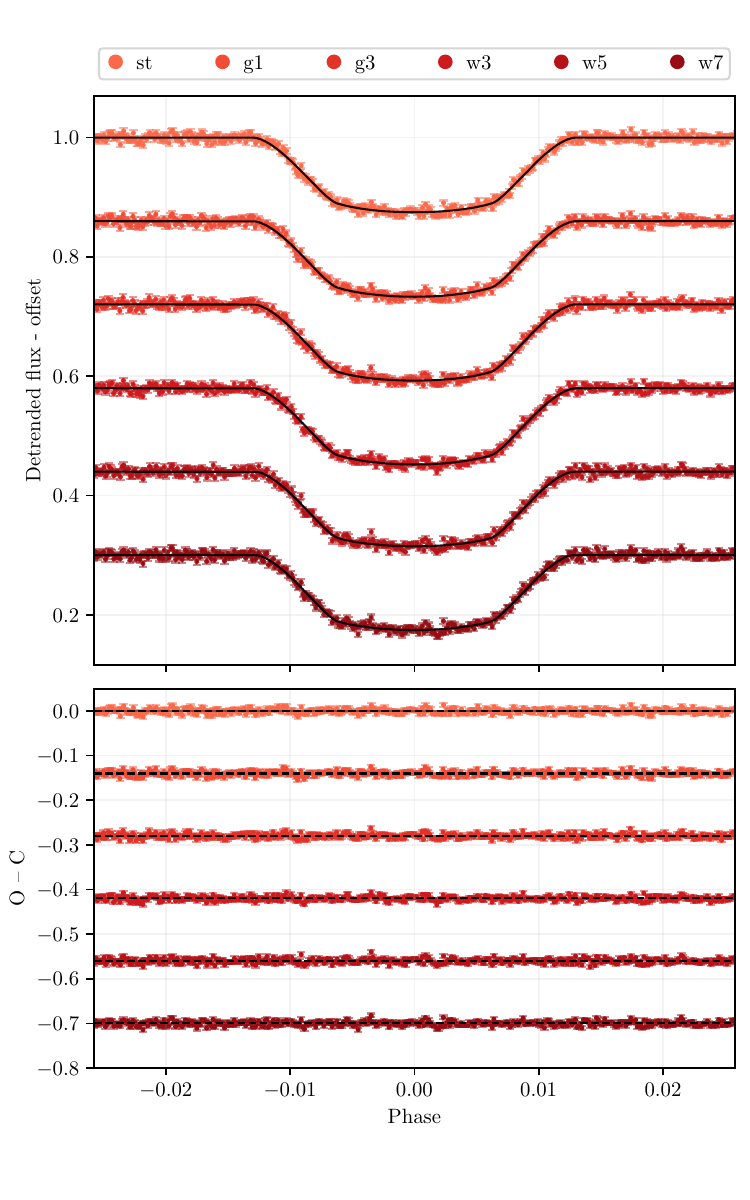}
    \caption{Similar to Fig.~\ref{fig: gp_lc} for i$^\prime$. The i$^\prime$ filter light curves are the least dispersed curves due to the number of counts reached for the target star being thousands higher than the dark current level of the image.} 
    \label{fig: ip_lc}
\end{figure}

\begin{figure*}
    \centering
    \includegraphics[width=17.8cm, trim={0.3cm 0.3cm 0.3cm 0.2cm},clip]{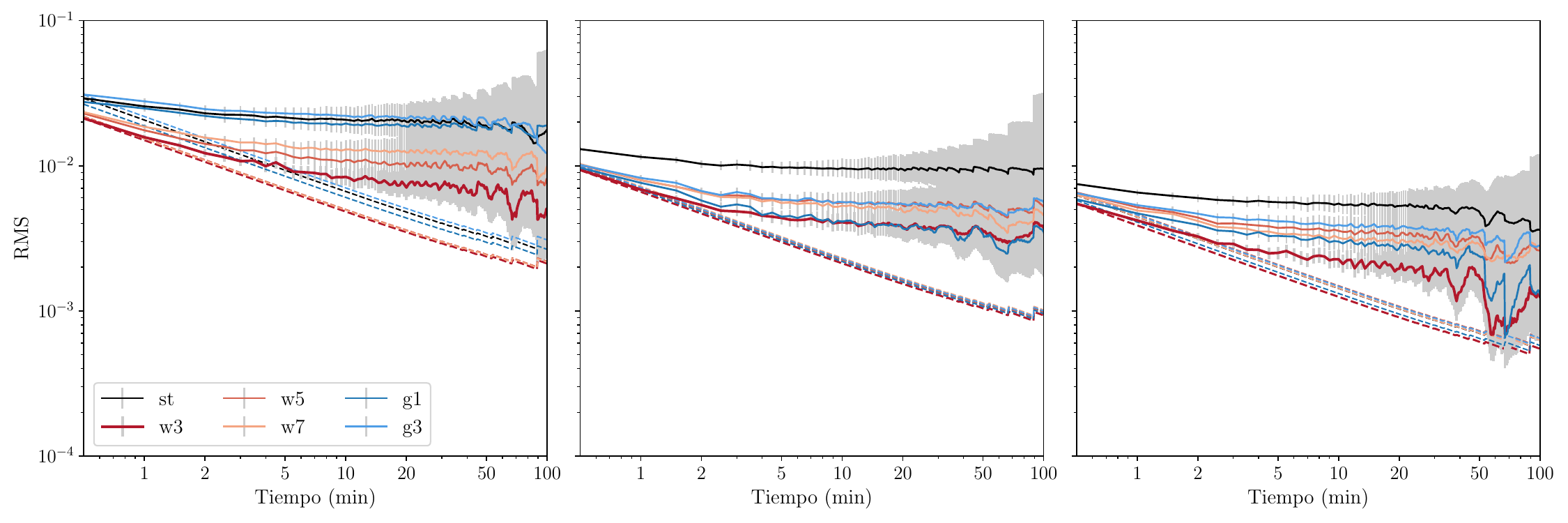}
    \caption{Time-averaging curves for all the reductions in g$^\prime$(left panel), r$^\prime$ (middle panel), i$^\prime$ (right panel). Standard reduction (st; black), 3$\times$3 median filter (w3; dark red), 5$\times$5 median filter (w5; red), 7$\times$7 median filter (w7; orange), gaussian kernel with 1$\sigma$ (g1; dark blue), gaussian kernel with 3$\sigma$ (g3; blue). The solid curves represent the time-averaging analysis, and the dashed curves represent the Gaussian noise contribution in the residuals. The closer the solid line is to the dashed line, the less correlated noise is in the light curve. The time-averaged curves for the reduction approaches illustrate how different reductions can mitigate red noise at different levels in light curves and how the redder bands have lower correlated noise.
    An interactive version of this plot is available at: \href{https://s-paez.github.io/opticam_lc/}{www.s-paez.github.io/opticam\_lc/}.}
    \label{fig:red-noise}
\end{figure*}

\begin{figure*}
    \centering
    \includegraphics[width=17.8cm, trim={5cm 3.4cm 5.3cm 6cm},clip]{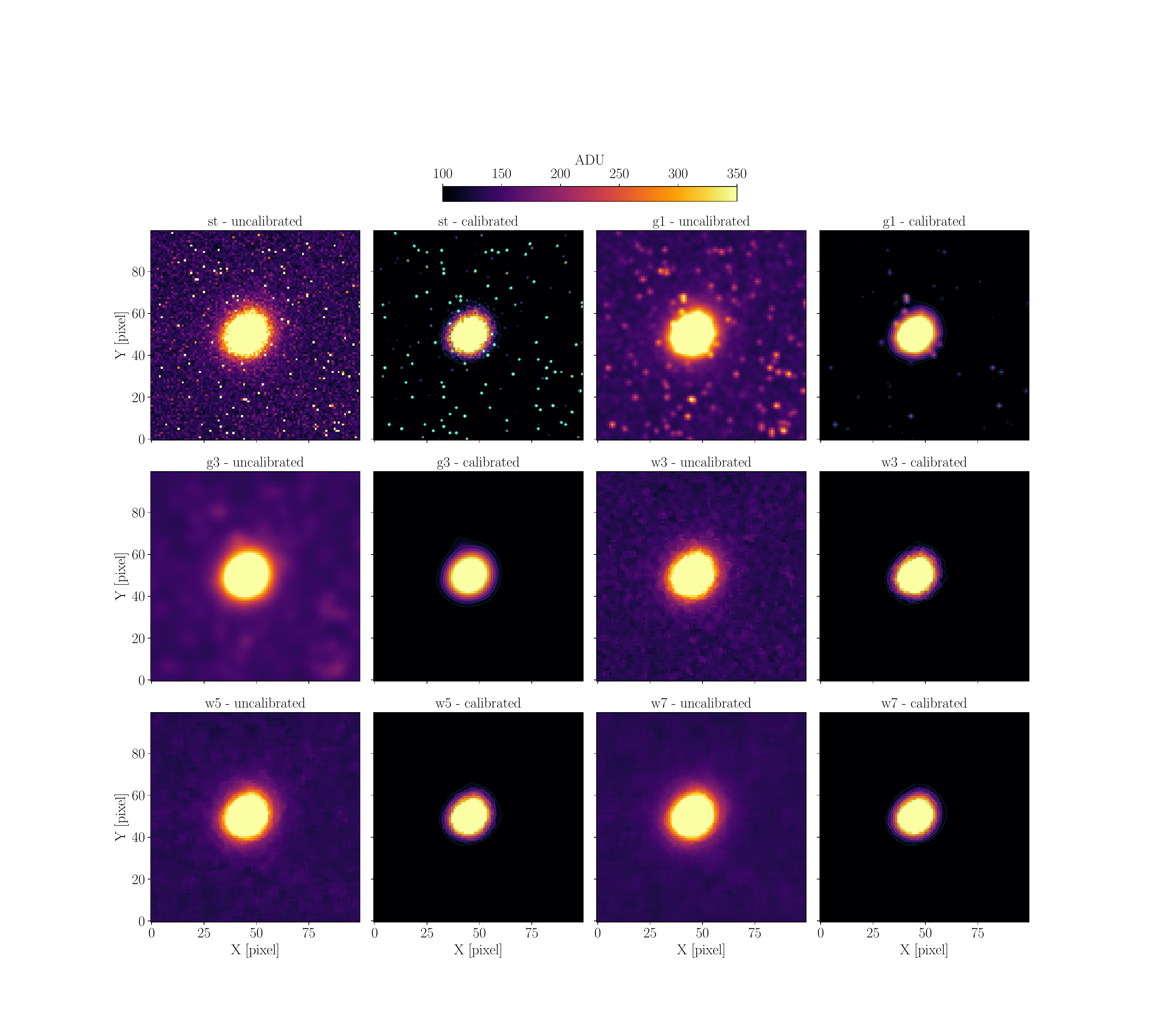}
    \caption{Image subframes of 13.9\arcsec$\times$13.9\arcsec centered around TOI-7149 in g$^\prime$ for the six reduction approaches, both before and after calibration. The first and third columns show the uncalibrated subframes. The second and fourth columns show the calibrated subframes. The calibrated subframes show cyan isocontours that reveal the star's final shape after each reduction. The panels illustrate how each method changes the appearance of the images and reveals different levels of residual warm-pixel across reductions. For example, \texttt{g1} reduction spreads the signal of a warm pixel and creates fake faint stars. Corresponding figures for r$^\prime$ and i$^\prime$ are provided in the appendices \ref{fig:all_results_rp} and \ref{fig:all_results_ip}, respectively.}

    \label{fig:all_results_gp}
\end{figure*}

We tested six reduction approaches of the OPTICAM transit of TOI-7149~b, fitting them with \texttt{juliet}, as described below in Section~\ref{sec:juliet}.
In Figures~\ref{fig: gp_lc}, \ref{fig: rp_lc}, and \ref{fig: ip_lc}, we show the g$^{\prime}$, r$^{\prime}$, and i$^{\prime}$ light curves, respectively,  obtained with the different reduction approaches, their fitted models, and their residuals. The interactive version of the figures\footnote{\href{https://s-paez.github.io/opticam_lc/}{https://s-paez.github.io/opticam\_lc/}} allows qualitative comparison of the light curves, models, and residuals for all three filters with all the reduction approaches explored in this paper. 
In Fig.~\ref{fig:red-noise}, we show the time-averaging curves for all the reductions in the three bands. This analysis involves measuring the RMS as a function of different temporal bin sizes in the residuals of light curves \citep{pont2006rednoise,cubillos2017oncorrelated}. The interactive version of Fig.~\ref{fig:red-noise} allows one to view and compare time-averaging curves for all filters and reduction approaches. 
Fig.~\ref{fig:all_results_gp} illustrates how each reduction approach has a different impact on the calibrated images and how the PSF is in each channel based on the isocountours. We can qualitatively see which methods leave more residual warm pixels in the subframes. To further quantify how the reduction approaches change the shape of the stellar signal, we provide the radial profiles for all reduction methods in the three bands in Appendix \ref{fig:radial_pro}. These profiles allow for a direct comparison of how each method alters the stellar peak counts. 

The apparent performance differences seen in Table~\ref{tab: metrics} across bands in the metrics are in part driven by the spectral energy distribution of TOI-7149. As an M dwarf, it is much brighter in the redder bands, so the stellar peak counts in r$^\prime$ and i$^\prime$ are far higher than in  g$^\prime$.
We describe the application and discuss the results method by method in the following sections.

\subsection{Standard reduction} \label{sec:st reduction}

The standard reduction uses dark and flat-field frames to correct for detector and optical system artifacts in science images \citep{howell2000handbook}. 
We use at least 20 dark frames per time series taken during the same observing campaign and with the same exposure time as the science observations.  Using AIJ \citep{collins2017astroimagej}, we create a master dark via a median combination of the darks.
This master dark is used to perform a dark subtraction on the flat-field and science images, to remove the signal generated by thermally induced electrons in the detector
\citep{howell2000handbook}.

Using AIJ, we create a master flat by combining the dark-corrected sky flats taken on the same night as the transit of TOI-7149~b (Section~\ref{sec:observations}) and normalize it to a median value of 1.0. 
This master flat is used to perform flat division on each science image to equalize the response of all pixels to incident radiation and to account for nonuniform illumination across the detector or within the optical system.

This method serves as a baseline for comparing the different reductions presented in the following subsections. Although the standard reduction for sCMOS detectors involves dark subtraction and flat-field correction, we found that this approach does not correct the warm pixels in the images (Fig.~\ref{fig:all_results_gp} and radial profiles in Fig.\ref{fig:radial_pro}). The science and dark frames are affected by warm pixels similarly, depending on exposure time, but those pixels are not the same.  Otherwise, dark subtraction would completely remove them.


Based on the metrics calculated for this reduction in Table~\ref{tab: metrics}, the standard reduction (\texttt{st}) produces light curves with a relatively high level of white and red noise, especially in the bluest band, where warm pixels are most problematic because of the lower S/N of the target star in this band. We therefore do not consider that the standard reduction is able to correct the OPTICAM data with sufficient precision for transiting planet modeling.

\subsection{Gaussian convolution kernel before standard reduction} \label{sec:gau_kernels}

We applied a Gaussian convolution kernel, a mathematical smoothing of the images, to both the raw science and calibration images. This pre-processing step was implemented using the \texttt{Gaussian2DKernel} and \texttt{convolve} functions within the \texttt{Astropy} convolution package \citep{astropy2022astro}. We tested two kernel widths defined by standard deviations of $\sigma$=1 pixel and $\sigma$=3 pixels. The first was chosen because it was the smallest possible kernel. The second was chosen as a test of a larger kernel. This method adjusts pixel values according to a Gaussian distribution, effectively reducing sharp variations while preserving overall structure \citep[e.g.,][]{aniano2011common, clark2018dustpedia}.  Fig.~\ref{fig:all_results_gp} illustrates the results obtained using the Gaussian convolution kernel fully specified by its standard deviation ($\sigma$) values of 1 and 3, respectively. Once every science and calibration frame is convolved with the Gaussian kernel, we apply the standard reduction.

In the case of a Gaussian kernel with $\sigma=1$, the subframe of the calibrated science frames reveals that this approach does not reduce the number of warm pixels in the image (Fig.~\ref{fig:all_results_gp}), although it lowers the peak counts of all pixels with the highest counts. In the image subframe, both before and after the calibration, the warm pixels with the highest counts and/or largest spread appear more significant than before applying the Gaussian-kernel convolution, and in some cases, they could create fake stars in the images. This behavior is particularly problematic for photometry, as it may introduce additional flux into the aperture if such artifacts occur near the target star and/or the comparison stars.

For the Gaussian kernel with $\sigma=3$, the distribution of pixels shows that this smoothing effectively removes the warm pixels effectively. However, it lowers the peak counts of the target star signal by $\sim$13\%.
Because the Gaussian convolution smooths out the signal, this change is a redistribution of the stellar signal over more pixels. A broader profile can require larger photometric apertures, increasing the contribution of the background, and it could degrade the effective signal-to-noise for faint targets. The change in the stellar profile caused by the Gaussian kernel pre-processing can be inspected in the radial profiles in Appendix~\ref{fig:radial_pro}.

Based on the metrics in the last and second-to-last lines in Table~\ref{tab: metrics}, the Gaussian-kernel reductions (\texttt{g1} and \texttt{g3}) yield RMS values that are comparable to the standard reduction in the three filters. However, they show the largest correlated noise in the resulting light curves in g$^\prime$. 
The $\chi^2_\nu$ values of \texttt{g3} are among the lowest for all methods, indicating an overall good match between the model and the uncertainties; however, this comes with poor red-noise mitigation in g$^\prime$ and r$^\prime$. Overall, the Gaussian kernels do not surpass the median-filter reductions, particularly \texttt{w3}, in either $\mathrm{RMS}_{10\,\rm min}$ or Bayesian evidence for the transit model. We can say that applying a Gaussian kernel to the data is disadvantageous because it blurs the single-pixel spikes characteristic of warm pixels, spreading their flux into adjacent pixels and effectively creating artifacts that resemble faint stars (Fig.\ref{fig:all_results_gp}). These artifacts can contaminate the photometric aperture, introducing systematic noise rather than removing it.

\begin{table*}
\centering
\caption{ Bayesian evidences, photometric precision, and fit metrics for the six reduction approaches. The metrics are defined as: \textit{$\ln{\mathcal{Z}}$}: Log-Bayesian evidence (higher values indicate statistically preferred model). \textit{RMS}: Root-mean-square scatter of the residuals (lower is better).  \textit{RMS$_{\rm 10\ min}$}: RMS of the residuals binned to 10 minutes, serving as a metric for correlated/red noise (lower is better). \textit{Median $\sigma$}: Median of the photometric errors. \textit{RMS/RMS$_{10\ \rm min}$}: Ratio of observed scatter to the 10 min binned scatter (higher values are preferred). $\chi_{\nu}^2$: Reduced chi-squared (values closer to 1 indicate consitency goodness-of-fit).}
\begin{tabular}{l|c|ccc|ccc|ccc|ccc|ccc}
\toprule
& $\ln \mathcal{Z}$ & \multicolumn{3}{c}{RMS [ppt]} & \multicolumn{3}{c}{RMS$_{10\ \rm min}$ [ppt]} & \multicolumn{3}{c}{Median $\sigma$ [ppt]} & \multicolumn{3}{c}{RMS/RMS$_{10\ \rm min}$} & \multicolumn{3}{c}{$\chi^2_\nu$}\\
 & & $g'$ & $r'$ & $i'$ & $g'$ & $r'$ & $i'$ & $g'$ & $r'$ & $i'$ & $g'$ & $r'$ & $i'$ & $g'$ & $r'$ & $i'$ \\
\midrule
\texttt{st} & $4464.34$ & 29.19 & 13.02 & 7.47 & 20.76 & 9.64 & 5.48 & \textbf{9.36} & \textbf{6.24} & 4.24 & 1.41 & 1.35 & 1.36 & 4.48 & 1.79 & 1.38 \\
\texttt{w3} & \textbf{5004.96} & \textbf{21.22} & \textbf{9.38} & \textbf{5.47} & \textbf{8.38} & \textbf{4.09} & \textbf{2.27} & 13.03 & 6.46 & \textbf{4.11} & \textbf{2.53} & 2.29 & \textbf{2.41} & 1.96 & 1.65 & 1.39 \\
\texttt{w5} & $4952.20$ & 22.90 & 10.14 & 6.54 & 10.96 & 5.58 & 3.56 & 13.13 & 6.46 & 4.39 & 2.09 & 1.82 & 1.84 & 2.01 & 1.60 & 1.41 \\
\texttt{w7} & $4983.93$ & 23.42 & 10.27 & 6.33 & 12.96 & 5.30 & 3.19 & 13.03 & 6.84 & 4.38 & 1.81 & 1.94 & 1.98 & \textbf{1.87} & 1.58 & 1.37 \\
\texttt{g1} & $4773.87$ & 27.51 & 9.68 & 5.87 & 19.56 & 4.12 & 3.00 & 10.58 & 6.64 & 4.30 & 1.41 & \textbf{2.35} & 1.96 & 3.25 & 1.58 & 1.27 \\
\texttt{g3} & $4951.57$ & 30.95 & 10.13 & 6.47 & 22.09 & 5.62 & 3.91 & 13.62 & 6.67 & 4.47 & 1.40 & 1.80 & 1.65 & 1.88 & \textbf{1.42} & \textbf{1.21} \\
\bottomrule
\end{tabular}
\label{tab: metrics}
\end{table*}

\subsection{Median filter before standard reduction} \label{sec: median filter}
 
Warm pixels behave as unusable data that must be replaced, turning mitigation of warm pixels into a missing-data problem  \citep[e.g.,][]{little2019statistical}. In astronomical image processing, median {filters have been widely used for this purpose, effectively mitigating noise and outliers while preserving signal structure \citep[e.g.,][]{Mitra2010efficient, Trubitsyn2021switching}.

This method performs pixel-by-pixel corrections by replacing each pixel with the median value of the nearest neighbors within a square window of a specified size. To evaluate its effectiveness, we tested window sizes of 3$\times$3, 5$\times$5, and 7$\times$7 pixels. The windows have widths that correspond to angular sizes of 0.417\arcsec, 0.695\arcsec, and 0.973\arcsec in Channel 1; 0.42\arcsec, 0.70\arcsec, and 0.98\arcsec in Channel 2; and 0.498\arcsec, 0.830\arcsec, and 1.166\arcsec in Channel 3. 
The first was chosen because it represents the nearest line of neighboring pixels to the warm pixel, while the other two also include the second and third lines, respectively.The edge pixels are treated via mirrored boundaries to avoid trimming the images. In the case of a 3$\times$3 median filter, only the last lines of pixels are affected. These lines are never used during photometry because even a slight shift in the star's centroid will cause the aperture to cross the image's limits. To maintain the same image format and size, we prefer not to trim the images. Avoiding using stars close to the border is always a good practice.}
We apply this median filter to the raw science and calibration images. We then apply the standard reduction process to the median-filtered images.

From the subframes shown in Fig.~\ref{fig:all_results_gp}, it is clear that the median filter effectively mitigates the warm pixel effect for all three window sizes, even before calibration. For the three window sizes, the peak counts of the target star are lowered by $\sim$15 counts for the 3$\times$3 and 5$\times$5 windows, corresponding to a $\sim$9\% decrease in peak counts. In the case of the 7$\times$7-pixel window, it lowered the peak counts by $\sim$20 counts ($\sim$13\% similar to the Gaussian kernel with $\sigma=3$) (radial profiles in Fig.~\ref{fig:radial_pro}). Furthermore, it successfully removes outliers without creating fake stars, unlike the Gaussian kernel corrections.

Among the three median-filter reductions (\texttt{w3}, \texttt{w5}, \texttt{w7}), the unbinned RMS values are very similar, indicating that all three median filters deliver comparable point-to-point precision and model fits. Major differences appear in the red noise mitigation. The 3$\times$3-pixel window reduction \texttt{w3} yields the smallest 10-min binned scatter in all three filters, which is significantly lower than 
for \texttt{w5} and \texttt{w7}. This is also evident in the time-averaging curves (Fig.~\ref{fig:red-noise}), where the \texttt{w3} curves in all three bands remain closest to the white-noise expectation (dashed curves) over the full range of bin sizes.

The pronounced effect of warm pixels in the $g^{\prime}$ light curve in metrics reported in Table~\ref{tab: metrics} is not due to a worse detector, but because the stellar signal has a lower signal-to-noise ratio than in r$^\prime$ and i$^\prime$. We thus suggest that to minimize the effect of warm pixels in the OPTICAM light curves, the peak counts of the target and comparison stars should ideally be above $\sim$1300 counts in 30s exposures, $\sim$1000 counts in 20s exposures, $\sim$700 in 15s exposures, and $\sim$400 in 10s exposures.
However, it is not always possible to achieve this, and it becomes more important to apply the 3$\times$3 median filter
when the signal-to-noise ratio of the star is close to 1.

The metrics show that the 3$\times$3 median-filter reduction provides the best performance between red-noise mitigation and overall model performance among all reduction approaches.

\section{Light curve fitting}\label{sec:juliet}

For each reduction, we simultaneously fit the g$^\prime$, r$^\prime$, and i$^\prime$ light curves using the same priors, parametrizations, and \texttt{dynesty} sampler configuration \citep{speagle2020dynesty}. The multiband model consists of a transit component and GPs for each band to detrend baseline variability and correlated noise. As the planetary nature of TOI-7149 b is confirmed \citep{Kanodia2025searching}, we assumed a non-chromatic transit and fit a single $R_p/ R_\star$ common to all bands.

\begin{figure}\centering
  \includegraphics[width=\columnwidth,trim={0.0cm 2.0cm 1.7cm 2.0cm},clip]{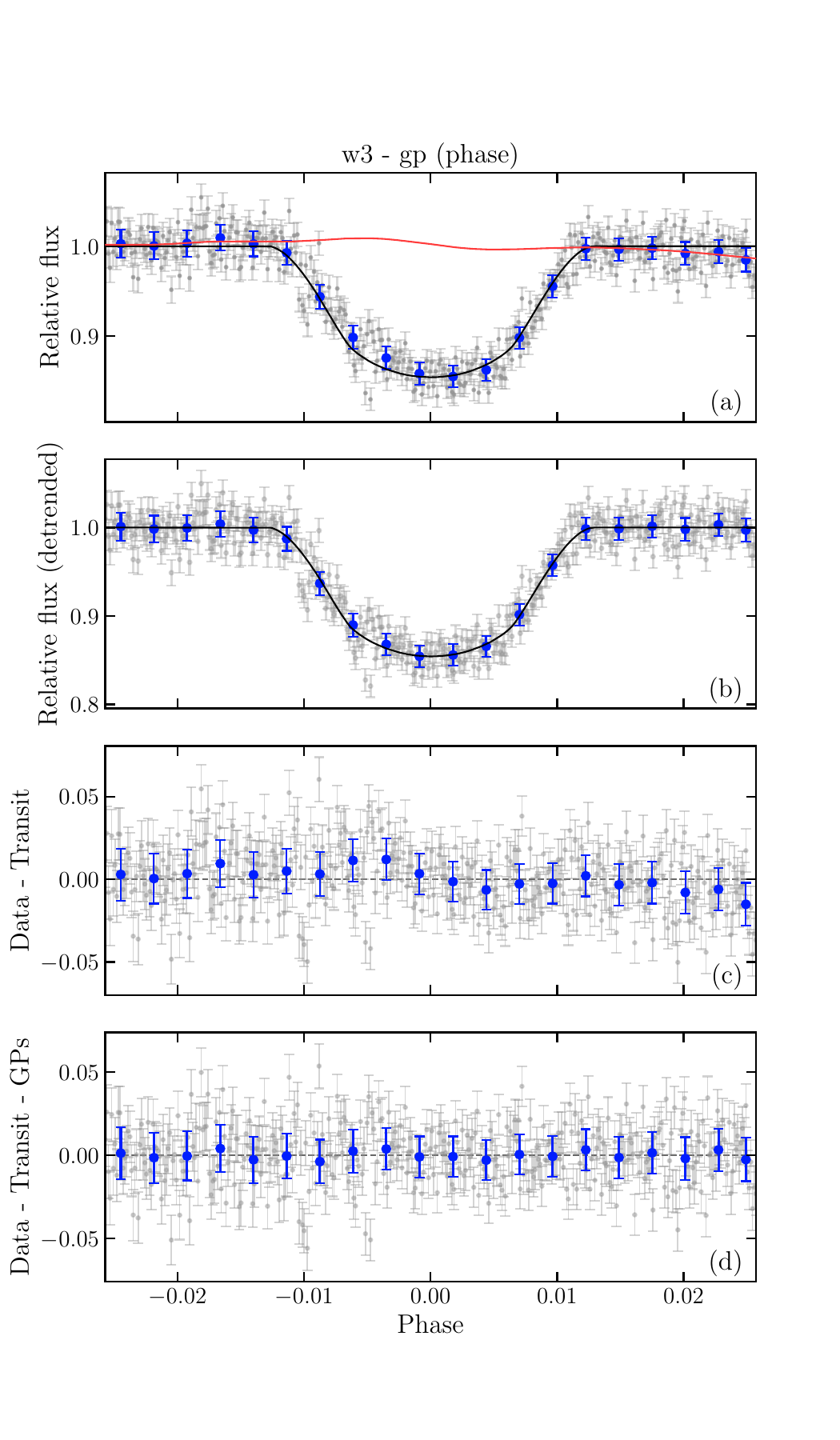}
    \caption{Example of the detrending performed on the g$^\prime$ transit light curve of  TOI-7149~b with the \texttt{w3} reduction. \textit{Top panel:} Light curve for \texttt{w3} reduction (gray dots with error bars) with 10-min bins (large blue dots with error bars); the transit model is plotted as a black solid line and the GPs model as a red solid line. \textit{Second panel:} Detrended light curve with GPs. \textit{Third panel:} Residual between relative flux and transit model without GPs detrending. \textit{Fourth panel:} Residuals between relative flux and transit model detrended with GPs. As the GPs correct for baseline variability, including correlated noise, we used the residuals without the GPs' detrending to perform the time-averaging analysis.}
    \label{fig: GPs_det}
\end{figure}

We incorporate detrending directly into the fit using the \texttt{celerite Matern kernel} \citep{Espinoza2019juliet} for each band, adopting the same priors for the GPs hyperparameters. This approach allowed us to fit the transit signal and the out-of-transit baseline simultaneously in the light curves (Fig.~\ref{fig: GPs_det}). The parameters common to all bands are $R_p/R_\star$ and the impact parameter $b$ (parametrized as $r_1$ and $r_2$ as suggested by \citet{Espinoza2018efficient}), the transit midpoint $t_0$, and the scaled semi-major axis $a/R_\star$. For each band, we fit the mean flux, the quadratic limb darkening coefficients (parametrized as $q_1$ and $q_2$ as suggested by \citet{kipping2013parametrizing}), an additional jitter term added in quadrature to the photometric uncertainties ($\sigma_w$), and the GP amplitude (GP$_\sigma$) and time scale (GP$_\rho$). 

\citet{Kanodia2025searching} derived the planet properties by fitting simultaneously the TESS light curve from the full-frame images, follow-up light curves, and high precision radial velocities. Given that we are fitting a single transit, we fixed the period to that reported by \citet{Kanodia2025searching}, and we used their transit midpoint with its uncertainty as the prior for our $T_0$. 
\citet{Kanodia2025searching} report an eccentricity of $e = 0.078^{+0.047}_{-0.048}$ with an argument of periastron $\omega = -0.286^{+0.692}_{-0.587}$ rad. To assess the effect of this small eccentricity on the 
derived light-curve dependent parameters, we fixed the eccentricity and the $\omega$ to their values and find that the $\ln\mathcal{Z}$ and the parameters are consistent to those with a circular orbit. Given the dataset we are working with, we consider that a circular orbit is a valid assumption for the orbit of the planet and adopt an eccentricity of zero for the rest of the analysis. 

We model the transit light curves using the static nested sampler \texttt{dynesty}, which provides both posterior samples and the Bayesian evidence for each model. We use 1000 live points to control the resolution with which the parameter space and the evidence are explored. We also apply a multi-ellipsoidal decomposition to approximate the high-likelihood region efficiently in the presence of degeneracies or multimodality. We use random-walk proposals within the ellipsoidal bounds that enclose the live points.
This procedure is well-suited to the moderately correlated posteriors typically encountered in models that include the transit and GPs \citep[e.g.][]{simpson2020marginalised, panwar2022new}.
Table~\ref{tab: fitting} summarizes the prior distributions adopted for the multiband fits. The first section lists the parameters common to all three bands, and the second lists the wavelength-dependent parameters. The references indicate the works that motivated our choice of values and/or distributions. We could not reproduce the light-curve dependent parameters reported by \citet{Kanodia2025searching}, because their solution combines radial velocity measurements, TESS photometry, and additional ground-based transit follow-up data that are unavailable. Their paper also does not report the priors used in the fit.
In our fits, the parameter with the largest offset relative to \citet{Kanodia2025searching} is $a/R_\star$. We attribute this difference to the higher photometric precision of our light curves compared to the light curves presented in the discovery paper. Our observations provide complete transit coverage with an extended out-of-transit baseline, which ensures robust normalization and precise characterization of the ingress and egress. A visual inspection of \citet{Kanodia2025searching}'s light curves reveals more dispersion than our data. Since $a/R_\star$ is constrained by the transit observation, we consider our derived value to be a more accurate refinement of the parameter.

\begin{table}
    \centering
    \caption{Priors for the multiband transit fitting}
    \begin{tabular}{lll}
    \hline
    Parameter      & Prior       & Ref.           \\ \hline
    P (days)       & Fixed (2.65206166) & A \\
    $T_0$ (BJD$_{TDB}$) & $\mathcal{U}$(2459703.596050, 0.00029)&A\\
    $r_1$  & $\mathcal{U}(0.0,1.0)$  & B           \\  
    $r_2$  & $\mathcal{U}(0.0,1.0)$  & B           \\
    $e$    & Fixed (0.0)& B\\
    a       & $\mathcal{U}(2.0,50.0)$       & B    \\ \hline
    mean flux  & $\mathcal{N}(1,0.01)$ & B         \\
    $q_1$, $q_2$ & $\mathcal{U}(0,1)$        & C \\ 
    $\sigma_w$ & $\log{\mathcal{U}} (0.0024,0.016)$& B\\
    GP$_\sigma$ & $\log{\mathcal{U}} (0.004,0.04)$& B\\
    GP$_\rho$ & $\log{\mathcal{U}} (0.05,0.15)$& B\\
    \hline
    \end{tabular}
    \newline
    {\footnotesize References: A=\citet{Kanodia2025searching}, B=\citet{Espinoza2019juliet}, C=\citet{kipping2013parametrizing} }
    \label{tab: fitting}
\end{table}

\begin{figure}
    \centering
    \includegraphics[width=\columnwidth]{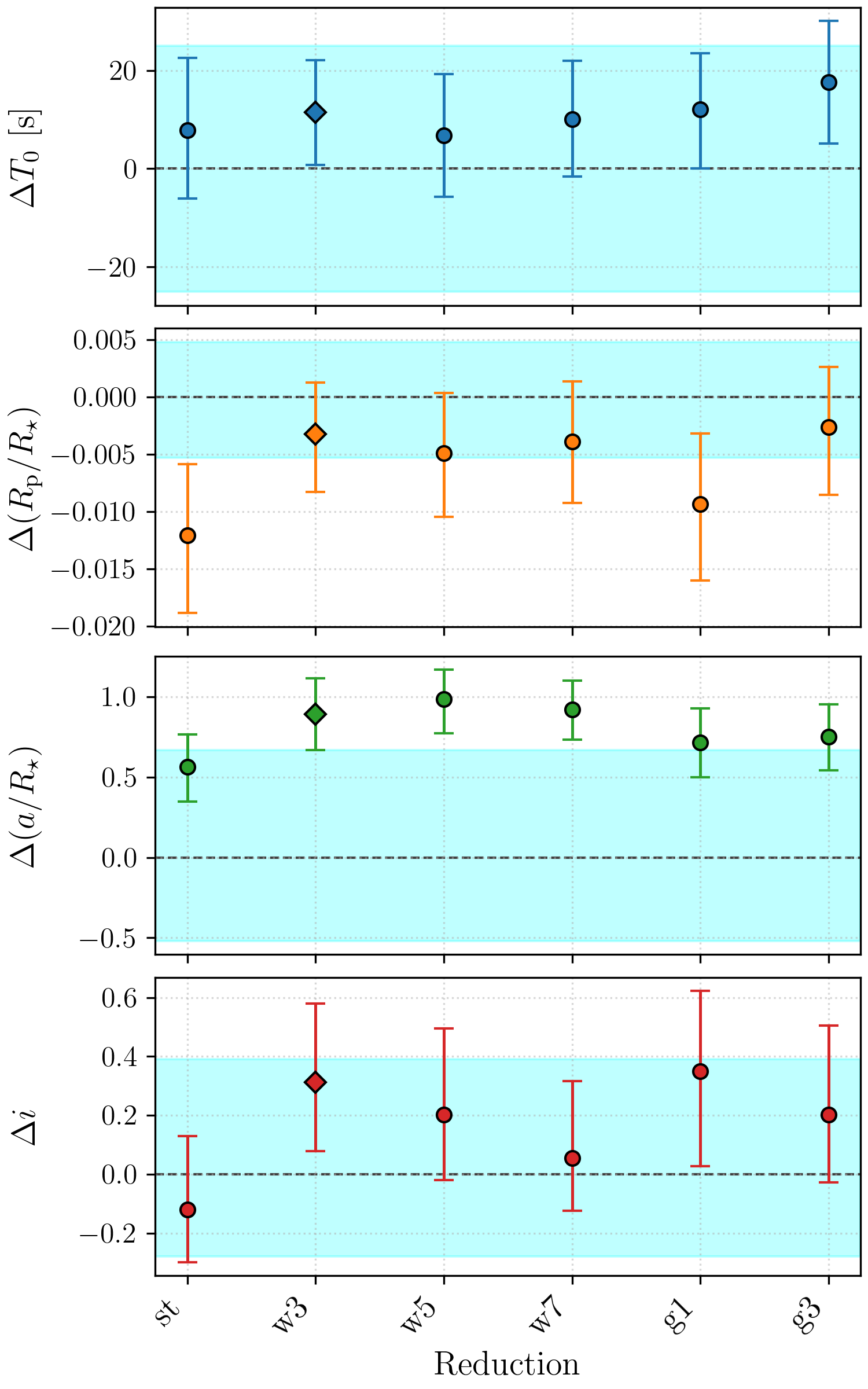}
    \caption{Differences in the planet parameters retrieved from our multiband fits for each reduction approach to those obtained by \citet{Kanodia2025searching} (cyan colored region). \textit{From top to bottom:} diference in transit midpoint ($T_0 [BJD_{\rm TDB}]$) in seconds, difference in scaled radius ($R_p/R_\star$), difference in scaled semi-major axis ($a/R_\star$), and difference orbital inclination ($i$ [deg]). The reductions are labeled as: \texttt{st} (standard reduction), \texttt{w3} (3$\times$3 median filter), \texttt{w5} and \texttt{w7} (5$\times$5- and 7$\times$7 median filters), \texttt{g1} and \texttt{g3} (1- and 3-$\sigma$ Gaussian kernels). Error bars denote the fitting uncertainties.}
    \label{fig: planet params}
\end{figure}

Table~\ref{tab:fit_comp} reports the median of posteriors for the parameters obtained from our multiband fits and the values reported by \citet{Kanodia2025searching}. While the wavelength-independent remain consistent across all reductions, the GPs hyperparameters and limb darkening coefficients exhibit larger dispersions or strong correlations. This behavior is expected because the single-transit fit lacks phase coverage, creating degeneracies between the correlated noise structure modeled by GPs and the stellar bright profile defined by limb darkening.

Regarding the results of the fitting, beyond the per-band metrics, the Bayesian evidences reported in Table~\ref{tab: metrics} reinforces the preference for \texttt{w3}. As discussed in Section~\ref{sec: median filter}, \texttt{w3} is decisively favoured over all other reductions. It yields $\Delta\ln\mathcal{Z}\gtrsim20$ relative to any alternative, a difference far exceeding the numerical uncertainty of the Bayesian evidence estimation which is of the order of $\sim$0.4. The standard reduction (\texttt{st}) has the lowest evidence and the largest $\mathrm{RMS}_{10\,\rm min}$ values, indicating that the light curves have significant correlated noise. The Gaussian-kernel models improve some aspects of the fit, but with a high correlated noise level in g$^\prime$ and r$^\prime$ and a relatively low $\ln\mathcal{Z}$. The 5$\times$5 and 7$\times$7 median filters achieve slightly different balances between white and red noise, but the \texttt{w3} has the best red-noise mitigation, with the lowest $\mathrm{RMS}_{10\,\rm min}$, and the highest Bayesian evidence. Globally, these metrics show that the 3$\times$3 median-filter reduction is the only approach among the ones we tested that simultaneously minimizes red noise and maximizes the probability that the transit model best describes the data. 

\begin{table*}
\centering
\caption{Posteriors of the parameters common to all bands in our multiband fits and the reference solution from \citet{Kanodia2025searching} (\texttt{K25}).}
\begin{tabular}{lllll}
\toprule
& $T_0$ ($\rm BJD_{TDB}$ & $R_{\rm p}/R_\star$ & $a/R_\star$ & $i$ (deg) \\
 &-2\,459\,700)
 \\
\midrule
\texttt{K25} & $3.596050\pm0.00029$ & $0.3334\pm0.0050$  & $15.52^{+0.67}_{-0.52}$  & $89.2^{+0.4}_{-0.3}$\\
\texttt{st} & $3.596140^{+0.00017}_{-0.00016}$ &$0.3213^{+0.0062}_{-0.0068}$& $16.08^{+0.20}_{-0.22}$ &$89.1\pm0.2$\\
\texttt{w3} & $3.596182\pm0.00012$ &$0.3301^{+0.0045}_{-0.0050}$& $16.41^{+0.21}_{-0.22}$ & $89.6^{+0.3}_{-0.2}$\\
\texttt{w5} & $3.596128\pm0.00014$ &$0.3285^{+0.0053}_{-0.0055}$& $16.50^{+0.18}_{-0.21}$ & $89.4^{+0.3}_{-0.2}$\\
\texttt{w7} & $3.596167\pm0.00014$ &$0.3295^{+0.0053}_{-0.0054}$& $16.44\pm0.18$ & $89.3^{+0.3}_{-0.2}$\\
\texttt{g1} & $3.596190^{+0.00013}_{-0.00014}$ &$0.3240^{+0.0062}_{-0.0067}$& $16.23^{+0.21}_{-0.22}$ & $89.6\pm0.3$\\
\texttt{g3} & $3.596254^{+0.00014}_{-0.00015}$& $0.3308^{+0.0053}_{-0.0059}$& $16.27^{+0.20}_{-0.21}$& $89.4^{+0.3}_{-0.2}$\\
\bottomrule
\end{tabular}
\label{tab:fit_comp}
\end{table*}

\section{Implementing the 3$\times$3 median filter reduction  with \profe\ and AstroImageJ} \label{sec: proposed method}
\begin{figure*}
    \centering
    \includegraphics[width=17cm, trim={0cm 2.2cm 43.5cm 0.5cm},clip]{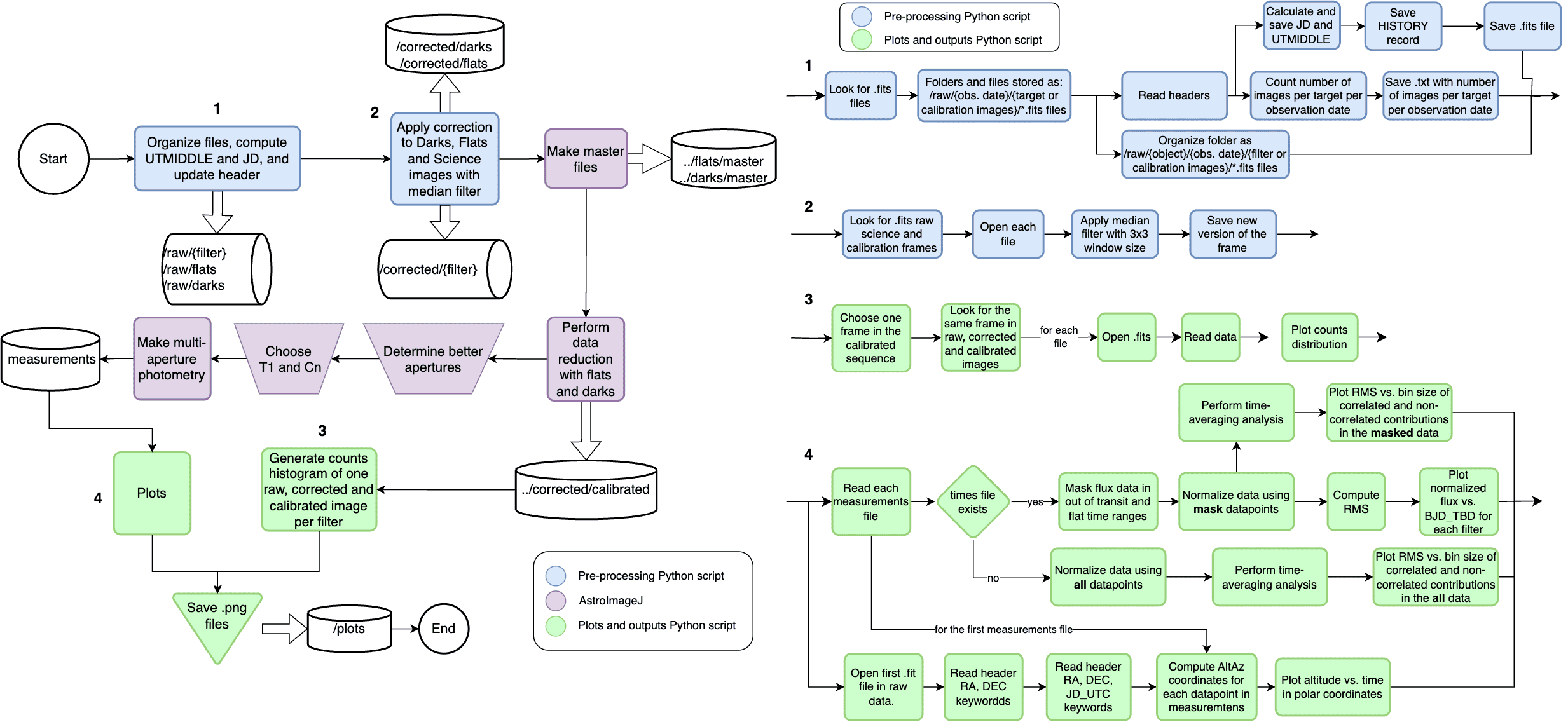}
    \caption{Data reduction flowchart.
    \textit{Blue boxes:} Steps in the process performed by the first part of the Python scripts (\profe).
    \textit{Rose boxes:} Steps in the process that are performed manually in the AIJ software. \textit{Green boxes}: Steps in the process performed by the second part of the Python scripts. Numbers reference sub-processes in each marked step and are detailed in Fig.~\ref{fig:subprocesses}. This pipeline is a well-integrated Python and AIJ tool that converts raw OPTICAM time series data into precise light curves of transiting exoplanets.}
    \label{fig:flowchart}
\end{figure*}

Following the support presented in the previous section, we propose the $3\times$3 median filter reduction as our preferred method to pre-process the OPTICAM data and obtain high-precision, multiband transit light curves.  
We suggest applying the 3$\times$3 median filter to all bands consistently, rather than selecting different reductions for different filters. As applying different pre-processing methods could introduce unquantifiable systematic errors.
Previous studies have demonstrated that data reduction and analysis homogeneity are crucial for precise planetary characterization \citep[e.g.,][]{2012ApJ...757..161T,ygmc2013homogeneous}. 
And \citet{Fulton2017radiusgap} showed that a homogeneous analysis was required to resolve the radius gap for small planets.

\profe\ is the set of Python modules that implement the automated parts of our OPTICAM reduction pipeline. The pipeline combines three elements: (i) pre-processing of raw and calibration frames with the 3$\times$3 median filter described in Section~\ref{sec: median filter} (handled by \profe), (ii) standard calibration and differential photometry performed in AIJ, and post-processing of the light curves and diagnostics (also handled by \profe). The \profe\ code therefore organises the data, applies the median-filter correction, updates metadata, and generates diagnostic products such as frame histograms, red-noise curves, and centroid movement plots. Due to the treatment of the headers, the three simultaneous channels are expected as inputs for the preprocessing and file formats, making a more generalised pipeline is beyond the scope of this paper. The creation of master darks and master flats, the calibration, and the differential photometry remain within AIJ. Figure~\ref{fig:flowchart} summarizes how these Python and AIJ components are connected within the overall pipeline.

In the pre-processing stage (blue boxes in Fig.~\ref{fig:flowchart}), \profe\ reads the raw science and calibration frames using the \texttt{Astropy FITS} module \citep{astropy2022astro}, organises them by target and observing night, and updates header information. It then applies the 3$\times$3 median filter correction to both science and calibration images using the \texttt{Scipy} median filter \citep{Virtanen2020SciPy}. 
The median-filter-corrected frames are written as new FITS files, which are subsequently handled with AIJ to create the master darks and master flats and to perform the standard calibration (master dark subtraction and master flat division). The differential photometry is also performed in AIJ, using fixed apertures and detector parameters appropriate for OPTICAM \citep{castro2024first}.
In the final stage of the workflow (green boxes in Fig.~\ref{fig:flowchart}), the \profe\ post-processing modules take the AIJ differential photometry tables and generate output products such as: normalized light curves, time-averaging (red-noise) curves, and centroid-movement plots for each target and observation night. 

A more detailed, step-by-step description of the file structure, configuration, and command-line interface of \profe\ and the operational details, such as configuration files, time-window selection, and command-line usage, are documented in the \profe\ repository\footnote{\href{https://github.com/s-paez/profe}{https://github.com/s-paez/profe}. }

\section{Summary}

OPTICAM sCMOS detectors exhibit warm pixels across both science and calibration frames, especially for exposures longer than 10 seconds. By using a sequence of dark frames at different exposure times, we are able to identify that the warm pixels have two main contributions: the dark current and higher bias level, and the salt and pepper noise and readout noise, which have an instrumental origin. 

The warm pixels affect the resulting light curves through white and correlated noise, and thus the retrieved planet parameters. To mitigate the impact of warm pixels on the light curves, we implemented two different pre-processing methods for the images before standard reduction: a Gaussian convolution kernel with a 1-$\sigma$ and one with a 3$\sigma$, and median filters with 3$\times$3-, 5$\times$5-, and 7$\times$7-pixel window sizes.

Our proposed method consists of applying a 3$\times$3-pixel window median filter to both science and calibration images before performing the standard reduction. This modification to the standard reduction removes warm-pixel artifacts with less degradation of the stellar signal, as measured by stellar peak counts.
Our proposed 3$\times$3 median-filter reduction achieves the strongest Bayesian evidence for the transit model and the best red-noise mitigation for all three bands. 

While in other metrics, different reductions achieve the best scores, none of them perform consistently well in all bands. For OPTICAM data, we consider that a small median filter window, like the \texttt{w3} approach, better reconstructs a given pixel value because larger median filters are more likely to include more than one warm pixel since the fraction of these can be up to $\sim$67\%. 

Finally, we introduced a reduction pipeline that combines \profe, a new open-source Python code, and AIJ, to implement our method optimized for transiting light curves acquired with OPTICAM at the OAN-SPM~2.1m telescope.

\section*{Acknowledgements}

Based upon observations carried out at the Observatorio Astron\'omico Nacional on the Sierra San Pedro M\'artir (OAN-SPM), Baja California, M\'exico. We gratefully thank the staff of the OAN-SPM for their valuable assistance during our multiple observing campaigns, and the OPTICAM team for their expert help with instrument operations. We are grateful to the anonymous reviewers for their valuable comments and constructive feedback, which helped to improve the manuscript. We thank Miguel Alarcón for his helpful suggestions regarding the analysis of the warm pixels' origin and behaviour. 
This research has been funded by UNAM PAPIIT IG-101224. S.P. acknowledges SECIHTI for the support through the Becas Nacionales de Posgrado program.

\section*{Data Availability}

The OPTICAM multiband light curves for TOI-7149 for all the reductions tested and reported in this study are available as online material.

\section*{Conflict of Interest}
Authors declare no conflict of interest.



\bibliographystyle{mnras}
\bibliography{example} 




\appendix

\section{Subframes before and after reduction for \MakeLowercase{r}$^\prime$ and \MakeLowercase{i}$^\prime$}

\begin{figure*}
    \centering
    \includegraphics[width=17.8cm, trim={5cm 3.4cm 5.3cm 6cm},clip]{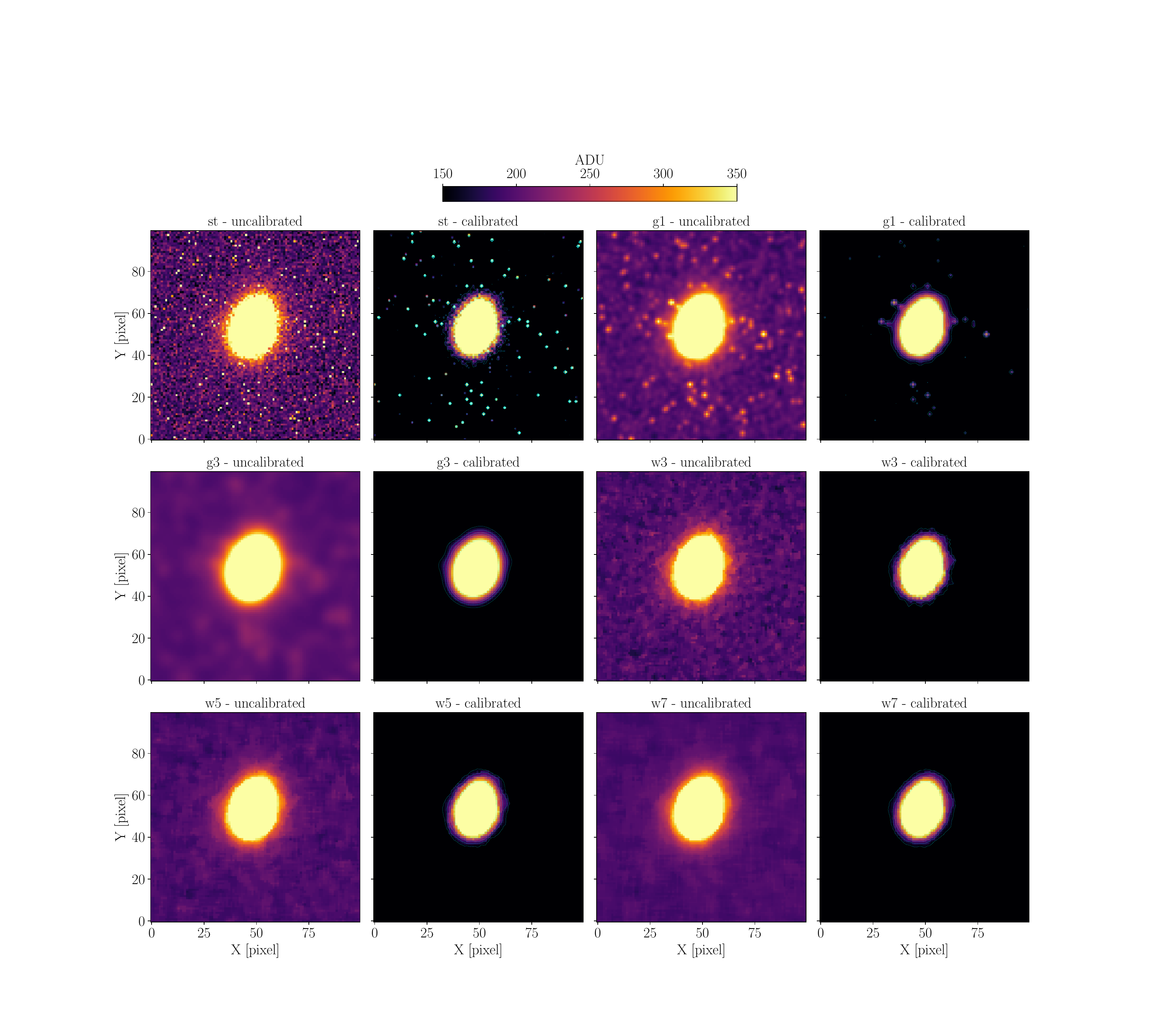}
    \caption{Image subframes of 14\arcsec$\times$14\arcsec centered around TOI-7149 in r$^\prime$ for the six reduction approaches, both before and after calibration. The first and third columns show the uncalibrated subframes. The second and fourth columns show the calibrated subframes. The calibrated subframes show cyan isocontours that reveal the star's final shape after each reduction. The panels illustrate how each method changes the appearance of the images and reveals different levels of residual warm-pixel across reductions.}
    \label{fig:all_results_rp}
\end{figure*}

\begin{figure*}
    \centering
    \includegraphics[width=17.8cm, trim={5cm 3.4cm 5.3cm 6cm},clip]{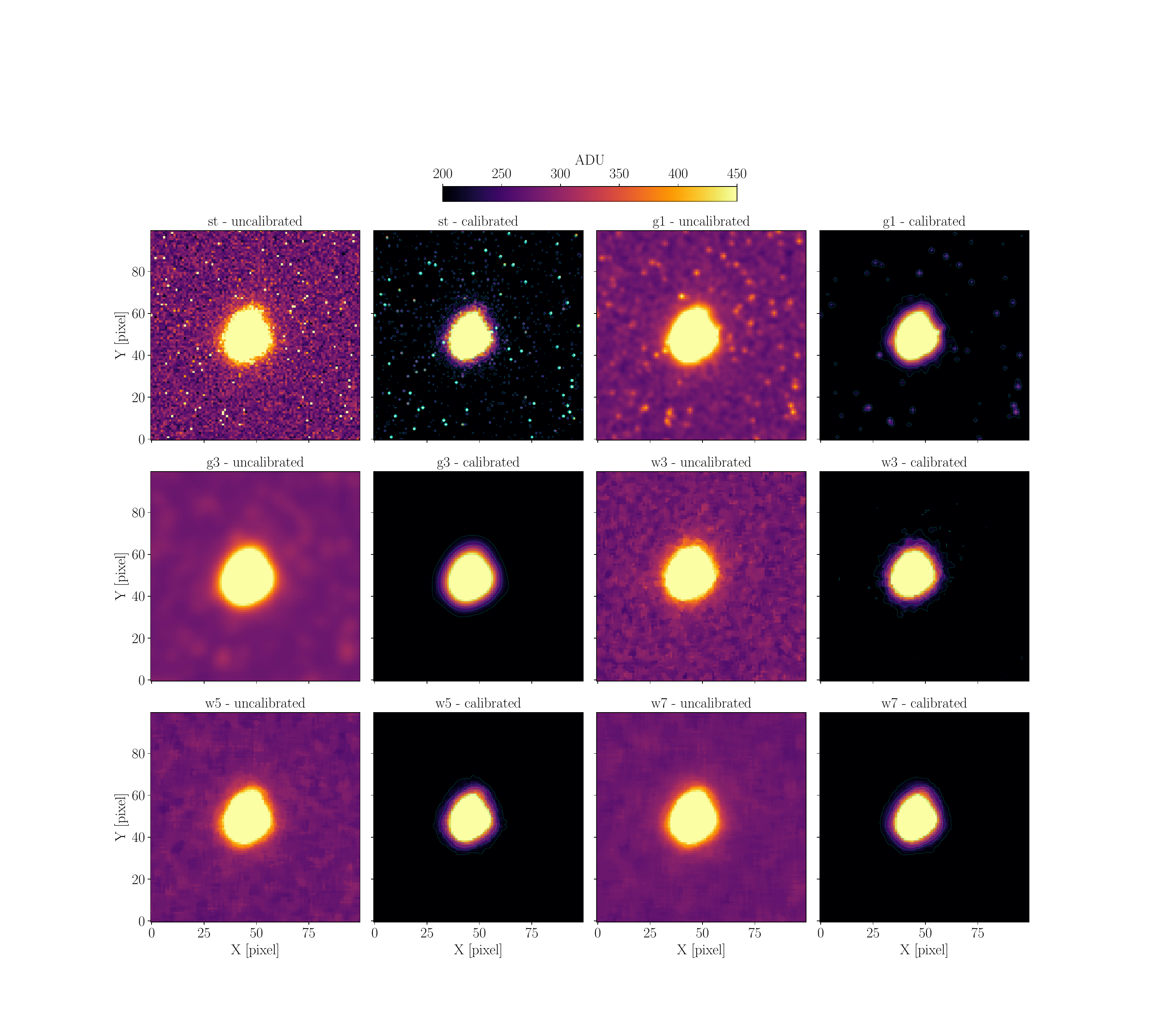}
    \caption{Image subframes of 16.6\arcsec$\times$16.6\arcsec centered around TOI-7149 in i$^\prime$ for the six reduction approaches, both before and after calibration. The first and third columns show the uncalibrated subframes. The second and fourth columns show the calibrated subframes. The calibrated subframes show cyan isocontours that reveal the star's final shape after each reduction. The panels illustrate how each method changes the appearance of the images and reveals different levels of residual warm-pixel across reductions.}
    \label{fig:all_results_ip}
\end{figure*}

Here we show the r$^\prime$ and i$^\prime$ summary panels (Figs.~\ref{fig:all_results_rp} and \ref{fig:all_results_ip}), which condense the subframes before and after calibration for each reduction method.

\section{Radial profiles}

\begin{figure*}
    \centering
    \includegraphics[width=\textwidth,trim={0.25cm 0.2cm 0.0cm 0.5cm},clip]{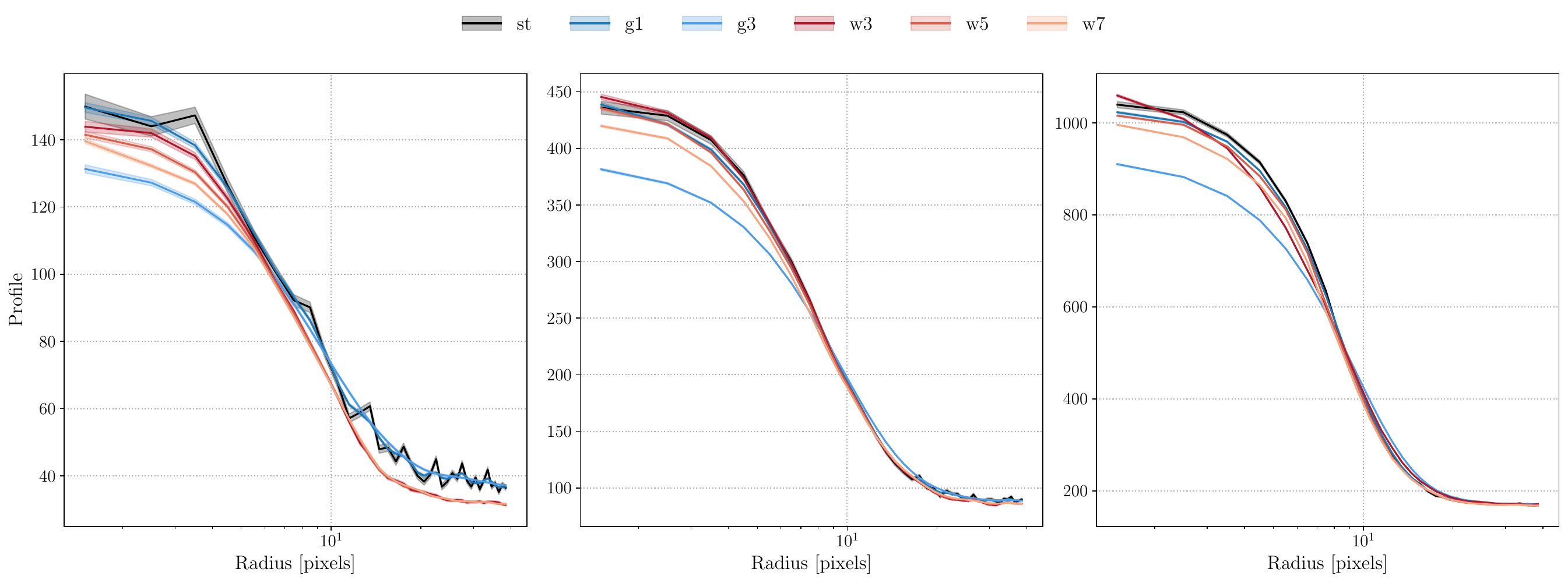}
    \caption{Radial profiles of TOI-7149 in g$^\prime$ (left column), r$^\prime$ (middle column), and i$^\prime$ (right column) for each reduction approach. \textit{From top to bottom}: \texttt{st}: standard reduction, \texttt{g1}: Gaussian kernel with $\sigma$=1 preprocessing, \texttt{g3}: Gaussian kernel with $\sigma$=3 preprocessing, \texttt{w3}: median filter with 3$\times$3-pixel window preprocessing, \texttt{w5}: median filter with 5$\times$5-pixel window preprocessing, and \texttt{w7}: median filter with 7$\times$7-pixel window preprocessing. These plots illustrate how each reduction approach mitigates warm pixels and reduces the star's peak counts.}
    \label{fig:radial_pro}
\end{figure*}

\section{Detailed processes in \profe's workflow}

\begin{figure*}
    \centering
    \includegraphics[width=17.8cm,trim={43.5cm 0cm 0cm 0cm},clip]{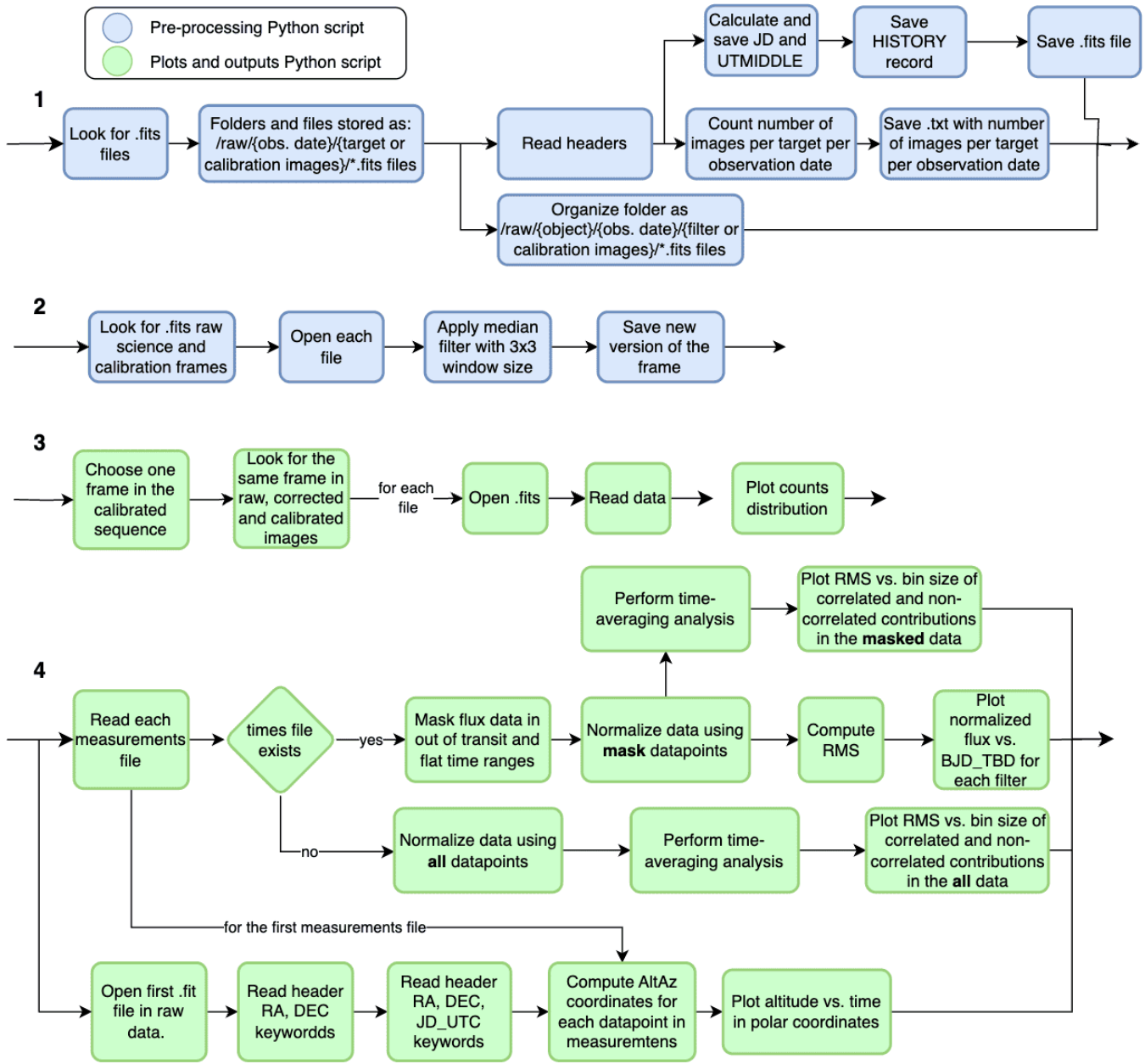}
    \caption{Sub-processes flowchart for the pipeline. \textit{Blue boxes}: Steps in the process performed by the first part of the Python script. \textit{Green boxes}: Steps in the process performed by the second part of the Python script. These flowcharts provide a detailed description of each step in the pipeline's Python scripts.}
    \label{fig:subprocesses}
\end{figure*}

Specific actions in the \profe's workflow are shown in Fig.~\ref{fig:subprocesses}. Each number in Fig.~\ref{fig:subprocesses}  refers to those steps in Fig.~\ref{fig:flowchart}. The combined information from Fig.~\ref{fig:flowchart} and Fig.~\ref{fig:subprocesses} presents the complete procedure of the data management, pre-processing, reduction, and post-processing.

\bsp	
\label{lastpage}
\end{document}